\documentclass[a4paper,twocolumn]{emulateapj}
\usepackage{amstext}
\usepackage{apjfonts}
\usepackage{amsmath}
\usepackage{color}
\usepackage{rotating}

\shortauthors{Naiman et al.}

\begin{document}

\title{External Mass Accumulation onto Core Potentials: Implications for Star Clusters, Galaxies and Galaxy Clusters}

\author{J. P. Naiman\altaffilmark{1}, E.
  Ramirez-Ruiz\altaffilmark{1} and D.~N.~C. Lin\altaffilmark{1}}
\altaffiltext{1}{Department of Astronomy and Astrophysics,  Theoretical Astrophysics Santa Cruz, University
  of California, Santa Cruz, CA 95064; jnaiman@ucolick.org}

\begin{abstract}
Accretion studies have been focused on the flow around bodies with point mass gravitational 
potentials, but few general results are available for non-point mass distributions.
Here, we  study the accretion flow onto non-divergent, core potentials  moving through a background 
medium. We use Plummer and Hernquist potentials as examples
to study gas accretion onto star clusters, dwarf and large galaxy halos and galaxy clusters in a variety of 
astrophysical environments.  The general 
conditions required for a core potential to collectively accrete large
quantities of gas from the external medium are derived using both simulations and analytic results. 
The consequences  of large mass accumulation in  galaxy nuclei, dwarf galaxies and star clusters are twofold.  First, if the gas cools effectively star formation can be triggered, generating new stellar members in the system. Second, if the collective potential of the system  is able to alter the ambient gas properties before the gas is accreted onto the individual core members,  the augmented  mass supply rates could significantly alter the state of the various accreting stellar populations and result in an enhanced  central black hole accretion luminosity.
\end{abstract}

\keywords{accretion, accretion disks, globular clusters:general, hydrodynamics, galaxies:general}

\section{Introduction}
The process by which stars gravitationally capture ambient matter is called accretion.
Calculations of the accretion flow onto stars are, in general,  very difficult.
First, one must determine the flow geometry; in general, if the gas possesses
intrinsic angular momentum, it will be multidimensional depending upon
the flow symmetry. Second, one must enumerate the dominant heating and cooling mechanisms
that characterize the accreting plasma. If the gas is optically thick to the emitted radiation, the net cooling and heating rates
will depend on the radiation field. Third, the effect of radiation pressure in holding back the flow must be
properly accounted for. Fourth, one must understand the flow boundary conditions, both at large distances and at the surface of the accretor.
Fifth, the possible role of magnetic fields in the plasma must be assessed. So, it is not surprising  that the problem of
gas accretion onto stars has been solved in only a few idealized cases.  In spite of the difficulties, significant progress has been
achieved in understanding accretion flows by means of such idealized solutions.

The problem of accretion of gas by  a star in relative motion with respect to the gas was first considered
by Hoyle and Lyttleton (1939) and later by Bondi and Hoyle (1944) (BHL accretion). The spherically symmetric case, when the
accreting mass is at rest with respect to the ambient gas, was first studied by Bondi (1952) and is commonly referred to as {\it Bondi} accretion.
This gives a reasonable  approximation to the real situation  of an isolated star accreting from the interstellar medium, provided
that the angular momentum, magnetic field strength and bulk motion of the interstellar gas with respect to the star can be neglected. For other types
of accretion flows, such as those in close binary systems, spherical symmetry is rarely a good approximation. Nonetheless, the spherical accretion problem
is of great relevance for the theory, as it introduces some key concepts which have much broader validity.  What is more, it is possible to give a fairly exact treatment,
allowing us to gain insight into more complicated  situations.

Accretion studies have been focused on the flow around bodies with point mass gravitational potentials, but few general results are available for non-point mass distributions. Mass models with tractable potentials may accurately  approximate the potentials of  galaxies, galaxy clusters and dense stellar systems.
In this paper we consider calculations of the gas flow  onto these
smooth core  potentials. We are interested in determining the amount of mass accumulation for core potentials with a
range of velocities relative to the ambient gas.  In \S2, we briefly recapitulate the relevant equations which describe the gas structure and
deduce analytically the conditions for generalized core potentials to accumulate gas collectively. Because the
gas structure is multi-dimensional in nature, we adopt a numerical approach to verify our analytic
solutions. We describe the numerical scheme and the range of model parameters in  \S3, and the results of these calculations in \S4.
These results are applied, in \S5, to study the accumulation 
of gas into dense stellar systems, galaxies and galaxy clusters moving  in a variety of environments.
In \S6, we summarize our results and
discuss their implications.

\section{Mass Accumulation in a Cluster Core Potential} \label{section:analytic} 

\subsection{Generalized Solutions}\label{subsection:general}

Much of the effort
herein will be dedicated to understanding the conditions by
which smooth core  potentials are able to accrete
surrounding gas with highly enhanced rates. In particular, we would like to
understand under which conditions a stable density enhancement will persist  within a
given
potential well and derive  a relation between the gas relative velocity, the local sound speed 
and the the central
velocity dispersion of the cluster potential,  which can be carried over quite generally to more complicated accretion flows.

To treat the problem mathematically we proceed as in \cite{doug}. We take spherical polar coordinates with origin at the center 
of the potential. In spherical symmetry, the fluid variables are independent of  angle  and the gas velocity $u$ has only a radial component.
For a steady flow, the continuity equation reduces to
\begin{equation}
r^2\rho u{\rm =constant}.
\label{mass}
\end{equation}
This constant is commonly denoted to as $\dot{M}$, and refers to a mass inflow (outflow) rate.
In the Euler equation the only contribution to the external force comes from gravity and this has only a radial component 
\begin{equation}
u {d u\over dr} +{1\over \rho} {d P \over dr} +  {d\phi \over dr} = 0. 
\label{euler}
\end{equation}
Hence, the term $({1/\rho})( {d P \over dr})$  is $(c_{\rm s}^2/\rho)(d\rho/dr)$, where $c_{\rm s}$ is the sound speed.
We replace the energy equation by the polytropic relation
\begin{equation}
P = K \rho^{\gamma},\;\;\;\;K{\rm = constant}.
\label{poly}
\end{equation}
In this case: $c_{\rm s}^2 = dP/d\rho = K \gamma \rho^{\gamma-1}$.
This allows us to approximately treat both adiabatic ($\gamma= 5/3$) and isothermal  ($\gamma= 1$)  gas flows. After the solution has been found, the adiabatic  or isothermal assumption should be justified by consideration of the particular radiative cooling and heating of the gas.   

The gravitational potential of dense stellar systems, galaxies and galaxy clusters can be  accurately described by  smooth core potentials with
\begin{equation}
\phi(r) = - \frac{\phi_0}{\left(r^\beta + \varpi_{\rm c}^\beta \right)^{\alpha}} = -\frac{\phi_0}{\zeta(r)},
\label{gpot}
\end{equation}
where $\varpi_{\rm c}$ is the radius of the core. The potential's core scale is given by 
\begin{equation}
\zeta(r) = \left(r^\beta + \varpi_{\rm c}^\beta \right)^{\alpha}
\end{equation}
Cores potentials described by  (\ref{gpot}) are 
non-divergent, so that $\phi(r \mapsto 0) = \rm{constant}$.  This property, as we will show, allows core potentials  to sustain subsonic inflow and collect ambient gas  into a quasistatic envelope.

An estimate of  the radial density distribution of collected ambient gas, $\rho(r)$,
follows from the application of the conservation laws for energy and momentum. From equation (\ref{euler}) we have  
\begin{equation}
\frac{u^2}{2} +  \int \frac{dp}{\rho}  + \phi = \frac{u^2}{2} + \frac{c_{\rm s}^2}{\gamma -1} + \phi = \rm{const}.
\label{denp}
\end{equation}
Equation (\ref{denp}) makes use of the polytropic equation of state, where    $dp = K \gamma \rho^{\gamma-1} d\rho$ 
and $c_{\rm s}^2 = \gamma p / \rho = K \gamma \rho^{\gamma-1}$.
Here, the limits $\phi(r \mapsto \infty) = 0$ and $u(r \mapsto \infty) = 0$ help determine the constant 
\begin{equation}
\frac{u^2}{2} + \frac{c_{\rm s}^2}{\gamma -1} + \phi  = \frac{c_\infty^2}{\gamma - 1}.
\label{denexp}
\end{equation}
Dividing  equation (\ref{denexp}) by $c_{\rm s}^2$ and making use of
$c_{\rm s}^2 / c_\infty^2 = \rho^{\gamma-1} / \rho_\infty^{\gamma-1}$, 
one recovers a modified form of the Bernoulli 
equation in \cite{bondi}
\begin{equation}
\frac{1}{2} \frac{u^2}{c_{\rm s}^2} + \frac{1}{\gamma-1} 
\left[ 1 - \left(\frac{\rho_\infty}{\rho}\right)^{\gamma-1} \right] = 
- \frac{\phi}{c_\infty^2} \left(\frac{\rho}{\rho_\infty}\right)^{\gamma-1}
\label{bern}
\end{equation}
The isothermal version of the  Bernoulli equation  can be derived from equation (\ref{denp}) using $dp = K d \rho$ and $c_{\rm s} = c_\infty$,
\begin{equation}
\frac{1}{2} \frac{u^2}{c_{\rm s}^2} + \log \rho + \frac{\phi}{c_\infty^2} = \log \rho_\infty.
\label{isobern}
\end{equation}

Together with knowledge of the  radial
velocity profile, equation (\ref{bern}) allows one to determine $\rho(r)$.
For  subsonic inflow $u^2(r)/c_{\rm s}^2 \ll 1$, the ambient gas collects into a  
a {\it quasistatic} envelope with a density distribution given by
\begin{equation}
\rho(r) \approx \rho_\infty \left[ 1 + \frac{(\gamma -1) \phi_0}
{c_\infty^2 \zeta(r) } \right]^{\frac{1}{\gamma-1}},
\label{corerho}
\end{equation}
where $\rho_\infty$ is the density of the ambient medium.   
For an isothermal 
equation of state, this result is simply
\begin{equation}
\rho(r) = \rho_\infty \exp\left[ \frac{\phi_0}{c_\infty^2 \zeta(r)}\right]
\label{isodens}
\end{equation}

In \S \ref{sub}, we  derive under what conditions we can safely ignore the $u(r)^2/c_{\rm s}^2$ term in equation (\ref{corerho}).

\subsubsection{Subsonic Flows in Core Potentials}\label{sub}
We now consider the validity of  the subsonic inflow assumption $| u(r)  | < c_{\rm s}$ used to derive  equation (\ref{corerho}) 
assuming subsonic flow at infinity, $ | u_\infty | < c_\infty$. In  spherically symmetry, 
the radial structure of the flow's velocity can be calculated by integrating ({\ref{euler}) with the help of
(\ref{poly}) and (\ref{mass}), so that
\begin{equation}
\frac{1}{2} \left( 1 - \frac{c_{\rm s}^2}{u^2} \right) \frac{d (u^2)}{dr} = {1 \over r}\left(2 c_{\rm s}^2 - r\frac{d \phi}{dr}\right)
\label{vel}
\end{equation}
At large distances from the potential the term $(2 c_{\rm s}^2 - r\frac{d \phi}{dr})$ on the right hand side of equation  (\ref{vel}) must be positive since $c_{\rm s}^2$ approaches  $c_\infty^2$ and $\phi \propto 1/r$ , while $r$ increases without limit.
 Given the that the inflow is initially subsonic, it will remain subsonic provided
\begin{equation}
c_{\rm s}^2 > \frac{r}{2}\frac{d \phi}{dr}
\label{cond}
\end{equation}
for all $r$ in the flow such that there is no transonic point.

Because $c_{\rm s}$ is a monotonically
decreasing function of $r$, equation (\ref{cond}) is satisfied provided 
\begin{equation}
c_\infty^2 > \frac{r}{2} \frac{d \phi}{dr}.
\label{cscond}
\end{equation}
Consequently, condition (\ref{cscond}) guarantees $| u(r) | <c_{\rm s}$ throughout  the  gas flow.  

Describing $\phi(r)$ as in equation (\ref{gpot}), condition (\ref{cscond}) can be rewritten as
\begin{equation}
c_\infty^2 > \frac{1}{2} \frac{\beta \alpha r^\beta \phi_0}{\left(r^\beta + \varpi_{\rm c}^\beta \right)^{\alpha+1}}.
\label{cscondcore} 
\end{equation}
The right hand side of equation (\ref{cscondcore}) has a maximum at 
\begin{equation}
r = \bigg( \frac{\varpi_{\rm c}}{\alpha} \bigg)^{1/\beta},
\end{equation}
so that condition  (\ref{cscond})  is satisfied for core potentials  if 
\begin{equation}
c_\infty^2 > \frac{\phi_0}{\varpi_{\rm c}^{\beta \alpha}} \frac{\beta}{2} \left(\frac{\alpha}{\alpha+1}\right)^{\alpha+1}. 
\label{subcond}
\end{equation}
Notice that  as $\varpi_{\rm c} \mapsto 0$, the ambient gas is unable to collect  into a quasistatic envelope and, as a result,
the density structure of the flow is no longer accurately described by   equation  (\ref{corerho}). 
In this limit, the potential essentially reduces to that of a point mass and the inflow rate 
can be derived following \cite{bondi}. 

 \subsubsection{Mass Accumulation in Non-Stationary Core Potentials}
The density structure of ambient gas collecting  into a core potential  in the case that
the gravitational structure is collecting gas isotropically is described by equation (\ref{corerho}).  If the potential 
is moving with some velocity with respect to the ambient medium, 
one expects the build up of a stable density enhancement to be inhibited.  A characterization of this impedance on the growth of the 
central density enhancement is derived here by comparing the
pressure support due to the collected ambient gas to that provided by the ram pressure. The radius at which these two 
pressures are equal defines a {\it characteristic radius}, which we denote here as $r_{\rm s}$.  
A schematic diagram illustrating the geometry  of the flow around non-stationary core potentials, together with a sketch of the expected
sound speed and density profiles of the   collected ambient material, is   given in Figure  \ref{fig:diagram}.

The characteristic radius, $r_{\rm s}$, is calculated here under the assumption that a stable density enhancement has been established, whose radial profile
 is well described by equation (\ref{corerho}). The potential, together with its collected ambient gas, is then assumed to move with a constant velocity, $u_{\rm const}\approx u_\infty = \mu_\infty c_\infty$. This provides a reasonable approximation provided that the flow variables across the shock region  are accurately described in terms of their limiting  values (i.e.  the incoming flow  is  not significantly altered by the gravitational potential and therefore the interior flow preserves its static solution). When this high velocity gas hits the stationary ambient medium, a bow shock forms. 
 In the frame moving with the shock, the density jump condition can be written 
 as
 \begin{equation}
\frac{\rho_\mu}{\rho_\infty} = \frac{(\gamma+1) \mu_\infty^2}{(\gamma+1) + (\gamma-1)(\mu_\infty^2 - 1)}
\label{pressure_ratio}
\end{equation}
where  $\rho_\mu = \rho(r)$, with $\rho(r)$ given by equation (\ref{corerho}). 
Equation (\ref{pressure_ratio}) assumes  the adiabatic index $\gamma$  is 
 the same for the post and pre-shock gas.  The jump conditions  across the shocked region yield
\begin{equation}
\zeta(r_{\rm s}) = \frac{\phi_0 (\gamma -1 )}{c_\infty^2} \left[ \left\{ \frac{(\gamma+1)\mu_\infty^2}
{2 + (\gamma-1)\mu_\infty^2}\right\}^{\gamma - 1} - 1 \right]^{-1}
\label{zeta}
\end{equation}
where we have used $\zeta(r_{\rm s}) = (r_{\rm s}^\beta + \varpi_{\rm c}^\beta)^\alpha$. 
These standoff distance define the {\textit{characteristic radius}} of the bow shock.
For $\phi_0 \propto M$ 
we have $\alpha \cdot \beta = 1$, so that 
 \begin{equation}
\zeta(r_{\rm s}= \varpi_{\rm c}) \equiv  2^{1/\alpha} \varpi_{\rm c}.
\label{rs}
\end{equation}  

 Equation (\ref{zeta}) can be rewritten for an isothermal 
equation of state as
\begin{equation}
\zeta = \frac{\phi_0}{c_\infty^2} {1 \over \log \mu_\infty^2}\;\;\;\;\;{\rm for}\;\;\;\;\; \gamma\rightarrow 1.
\end{equation}

In general, core potentials moving at high velocities have density enhancements and 
characteristic radii that decrease with increasing velocity.  At sufficiently 
high Mach numbers, one expects the incoming flow's ram pressure to overcome the 
central  pressure due to the collected ambient gas and to penetrate into the potential's core.  
The transition between a  stable density enhancement and an 
unsupported, supersonic inflow occurs roughly when $r_{\rm s} \approx \varpi_{\rm c}$. When  $r_{\rm s} \leq \varpi_{\rm c}$ , the ability of the core to sustain
 a stable density 
enhancement is lost.  For this reason, we expect rapidly  moving potentials to 
have smaller density enhancements than those predicted by equation (\ref{corerho}).  
In these cases our formalism inherently overestimates the internal pressure support against the inflowing ambient material, which 
in turn overestimates the characteristic radius. The detailed 
behavior of the characteristic radius with adiabatic index, Mach number 
and gravitational potential  type  will be explored in \S \ref{simulations} using numerical simulations.
 In the following numerical models, we will approximate isothermal equations of 
state by setting $\gamma= \, 1.01$.  For consistent comparison, we will use equations 
(\ref{corerho}) and (\ref{rs}) 
with $\gamma = \, 1.01$ in all plots unless otherwise noted.

\subsection{Specific Core Potentials}
In this paper we focus our attention  on the ability of  stellar clusters, galaxies, and galaxy 
clusters to  collect ambient material. 
Stellar clusters can be accurately described by Plummer potentials \citep{bruns,pf}
\begin{equation}
\phi(r) = \frac{G M_{\rm p}}{(r^2 + r_c^2)^{1/2}},
\label{plummer}
\end{equation}
where $M_{\rm p}$ is the mass of the potential and $r_{\rm c}$ is the 
commonly referred to as {\it core radius}.  In this case, $\phi_0 = G M_{\rm p}$, $\varpi_{\rm c} = r_{\rm c}$,
$\beta = 2$, and $\alpha = 1/2$,  so that equation (\ref{corerho}) can be rewritten as
\begin{equation}
\rho(r) = \rho_\infty \left[ 1 + \frac{G M_{\rm p} (\gamma -1)}{c_\infty^2 \left(r^2 +
r_{\rm c}^2\right)^{1/2}} \right]^{\frac{1}{\gamma-1}}
\label{eq:dcl}
\end{equation}

Dark matter profiles of galaxies and galactic clusters are, on the other hand, commonly modeled by  \citet{hernquist} potentials
\begin{equation}
\phi(r) = {G M_{\rm h} \over r + a},
\label{hernquist}
\end{equation}
where $M_{\rm h}$ is the mass of the potential and $a$ is the 
scale length.  
In these cases, $\phi_0 = G M_{\rm h}$, $\varpi_c = a$, 
$\beta = 1$, and $\alpha = 1$, so that equation (\ref{corerho}) becomes
\begin{equation}
\rho(r) = \rho_\infty \left[1 + \frac{GM_{\rm h}(\gamma-1) }{c_\infty^2 (r + a)}
\right]^{\frac{1}{\gamma-1}}
\label{eq:dhe}
\end{equation}
In what follows, these two types of potentials are numerically modeled moving through 
ambient  gas at rest to and compare to our analytic treatment.

\section{Simulations of Mass Accumulation in Core Potentials}\label{simulations}

\subsection{Numerical Method and Initial Model}
To examine the ability of core potentials  to collect ambient gas  into a quasistatic envelope,
we simulate core potentials moving through ambient gas with FLASH, a parallel, adaptive mesh
refinement hydrodynamics code. This scheme, and tests of the code are
described in \citet{fryxell}.

Common to all calculations is the placement  of  inflow boundaries to one side of our computational domain in order to simulate the
motion of the core potential through the external medium.  As a result, ambient gas moves to the right of our grid at a supersonic velocity.  When this high velocity gas hits the stationary potential, a bow shock forms. All gravitational potentials  are
modeled here using  a Plummer (\ref{plummer}) or a  Hernquist (\ref{hernquist}) functional form.

Simulations   start with a uniform background density and run until a steady
density enhancement is established inside the core's potential. This usually takes
about 10-100 sound crossing times \citep{us}.
Several models were run longer to
test convergence and density enhancements were found to change only
slightly with longer run times.  We further tested convergence of our
models for several resolutions and domain sizes.  All tests produced
similar density enhancements to those shown here.  After hundreds of
sound crossing times, the flow is relatively stable, and does not
exhibit the {\it flip-flop} instability seen in two dimensional
simulations \citep{2009ApJ...700...95B}.

The effects of self gravity of the gas are ignored. This is adequate
for most of our models, for which the accumulated mass is significantly less than the
mass responsible for the potential.   To improve the
controlled nature of the models, and to facilitate comparison with our analytic
estimates, we do not explicitly include radiative heating or cooling.  As
with our analytic estimates, we assume a polytropic equation of state $P \propto \rho^\gamma$.
The effects of radiative equilibrium are approximated by having the gas evolve with
an adiabatic constant $\gamma= 1.01$ (giving nearly isothermal
behavior), while  inefficient cooling is model here by having the gas evolve with
$\gamma= 5/3$.

\subsection{General Properties of the Gas Flow}\label{general}
The classical Bondi treatment for non-divergent potentials, representative of
star cluster and galactic gravitational potentials,  is only
a fair approximation far from the core when $GM_{\rm c}/\varpi_{\rm c}
\gg c_\infty^{2}+u_\infty^{2}$. When $GM_{\rm c}/\varpi_{\rm c} \lesssim c_\infty^{2}+u_\infty^{2}$, the collective potential alters the local gas
properties, allowing the core  to sustain subsonic mass inflow and, as a result, collect ambient gas  into a quasistatic envelope.

Figure \ref{fig:comp} shows the resulting density  structures of the collected  ambient gas for a point mass and
two different core potentials.  To facilitate comparison,
the Plummer and Hernquist potentials depicted in Figure \ref{fig:comp} have been constructed to yield the same circular velocity peak radius: $r_{\rm v}$.
This requirement can be rewritten as
\begin{equation}
a= \sqrt{2} r_{\rm c},
\label{eq:arc}
\end{equation}
which, by demanding the mass within this characteristic radial scale to be the same, gives
\begin{equation}
M_{\rm h} = {8\over 3} \sqrt{2\over3} M_{\rm p}.
\label{mhmp}
\end{equation}
Hence,  equations (\ref{eq:dcl}) and (\ref{eq:dhe}) give
\begin{equation}
{\rho_{\rm h} (r=0)\over \rho_{\rm p} (r=0)}\approx \left({8 \over 3\sqrt{3}}\right)^{1\over \gamma-1},
\end{equation}
where $\rho_{\rm max} \equiv \rho(r=0)\gg \rho_\infty$.

The flow pattern around a core potential  is
multi-dimensional and complex. In the frame of
the potential, the gas streamlines are bent towards the cluster
center. Some shall intersect the center, while others converge along a
line behind it. The convergence speed of the gas determines the
reduction in its velocity relative to the potential due to shocks, and
therefore whether or not the gas is accumulated in the core.  As seen in Figure \ref{fig:comp},  the flow structure  within the cluster markedly differ from the classical Bondi flow.

The properties of the core potential have  a decisive effect on the radial profiles of the collected ambient gas. For $r \ll \varpi_{\rm c}$ and $\rho_{\rm max}\gg \rho_\infty$, equation (\ref{corerho})
gives
\begin{equation}
\frac{\rho(r)}{\rho_{\rm max}} \approx \left[ 1 - \alpha \left(r/\varpi_{\rm c}\right)^\beta\right]^{\frac{1}{\gamma-1}}
\label{rhodiff}
\end{equation}
This shows  that gas  collected in potentials with steep core profiles have, as expected, sharper central
density profiles.    From equations (\ref{rhodiff}) and (\ref{eq:arc}), it follows that
 $\rho/\rho_{\rm max} \propto [1- (r/\sqrt{2}r_{\rm c})]^{{1 \over \gamma-1}}$  
 and $\rho/\rho_{\rm max} \propto [1- 2 (r/r_{\rm c})^2]^{{1 \over \gamma-1}}$
for  Hernquist and  Plummer potentials, respectively. 
Figure \ref{fig:comp} shows that the the steeper Hernquist interior potential 
 produces gas accumulation which 
is more centrally confined within its core than that in the Plummer potential.

At $r \gg \varpi_{\rm c}$, the divergence of the gravitational potential naturally leads to a loss of pressure support   against the incoming material.
As a result, the characteristic radius, $r_{\rm s}$, will be smaller for less extended (or less massive) core  potentials.
 Inserting equation (\ref{mhmp}) into equation ({\ref{zeta}) and making use
of equation ({\ref{rs}), one finds
\begin{equation}
\frac{\zeta(r_{\rm s} = \varpi_{\rm c})_{\rm h}}{\zeta(r_{\rm s} = \varpi_{\rm c})_{\rm p}} 
= \frac{8}{3 \sqrt{3}}
\label{sizecomp}
\end{equation}
for Hernquist and Plummer potentials yielding the same circular velocity peak radius,
$\gamma$, $\mu_\infty$ and $c_\infty$.
Under these conditions,  the 
 ambient gas collected into a  Hernquist potential at $r\gg \varpi_{\rm c}$ will be comparatively denser and thus  lead  to an increase in  pressure support  against the incoming flow. 
This is mainly due  to the comparatively larger mass  (see equations (\ref{mhmp}) and (\ref{zeta})) together with the  less divergent potential at large radii.  This enhanced support forces the flow to move around the central
density enhancement, thus leading  to a widening of the standing bow shock.
This is evident   in Figure \ref{fig:bows}, which  plots the  bow shock patterns  produced  around a  point mass \citep{ruffert}, Plummer and Hernquist potential set in motion through an initially uniform medium.

\subsection{Mach Number Dependence}

In line with the conventional treatment, potential cores moving with respect to the
ambient  medium at increasing supersonic velocities will have
density enhancements that are progressively smaller and significantly
more offset from the cluster's center.   Figure \ref{fig:machc} shows
the resulting density and velocity  flow structures generated  around a  point mass, Plummer and Hernquist potential
moving  through an initially uniform medium with $\mu_\infty\geq 1$.   The differences in the flow structure  are most notable when comparing the flow velocity structures depicted in Figure  \ref{fig:machc}, which show that the central density enhancements in core potentials are quasi-static, while those for the point mass are not.
While there are regions of  inflowing gas around core potentials, primarily
behind the bow shock and downstream from the core's center, there are
large central regions that experience no net inward gas flow.

A stable density enhancement is observed to persist  as long as $\mu_\infty\lesssim 3$ (Figure \ref{fig:machc}).
For larger relative velocities,  the central gas pressure support is overwhelmed by the
ram pressure of the incoming  gas and, as a result, the bow shock begins to
detach from the potential's core. The transition from a bow to a tail shock
takes place at around $\mu_\infty \approx 3$. Note that in the case of a point mass potential with an arbitrarily
small sink radius, the gravitational
forces will always overwhelm the ram pressure  at a finite radius. In this case, there is no transition  to a tail shock \citep{wolfson}.
However, for core potentials (and point mass potentials with a
finite sink size), the gravitational force approaches a constant value
near the potential's center, thus allowing the formation of  a tail shock.

As shown in Figure~\ref{fig:wanal}, the  radial density profiles of the accumulated ambient gas,
within a core potential  differ  from the classical Bondi profiles, and, for
low Mach numbers, are accurately  described by the quasistatic
solutions derived in \S \ref{section:analytic}.  These
solutions fail to provide a good description of the density of the inflowing gas
for $\mu_\infty \gg 1$. This is because these solutions  were derived assuming a static ambient gas.
The decrease  of the characteristic radius and of  the collected  ambient gas density with increasing $\mu_\infty$  are clearly visible
in the simulations. This decrease is connected with an increase of the ram pressure of the incoming material which, in turn, helps overcome the
central pressure of the collected ambient gas.

\subsection{Dependence on the Polytropic Index $\gamma$} \label{section:gammadep}
The determination of the internal structure of gas  flowing around gravitational potentials
depends upon detailed knowledge of the equation of state.  In \S \ref{section:analytic}, we have derived
analytical solutions for the internal pressure and density structure of the flow
in the cases in which the pressure depends solely upon the density through an equation
of state of the form  $P=K \rho^\gamma$.

The dependence of the  overall gas density structure on the polytropic index
$\gamma$ is illustrated  in Figures \ref{fig:gammas} and
\ref{fig:gammasb}.   By comparing  the flow structures  in Figures   \ref{fig:gammas} and
\ref{fig:gammasb}, one sees that  ambient gas  is in fact collected  much more gradually  when its pressure support  increases adiabatically ($\gamma = 5/3$). From equation (\ref{corerho}), it follows that $P\propto \zeta^{-\gamma \over \gamma-1}$. Thus,  gas accumulated in potentials with higher adiabatic indexes will be less centrally  confined and, as a result,
will be less effective at protecting  the collected  gas  from being displaced by the inflowing material.
 By contrast,  gas collected isothermally ($\gamma \approx 1$) into a  core potential  will be comparatively denser and thus  lead  to a significant  increase in  pressure support at small radii.
This  is consistent with the behavior observed in Figure  \ref{fig:gammas}, where decreasing $\gamma$ moves the shock closer to the potential's core. Despite the initially  less extended bow shock,  the enhanced support  provided by  more compressible flows  at small radii delays the transition  to a tail shock as the  relative velocity increases and, as a result, the bow shock radius decreases. 

\subsection{General Conditions Required for Large Density Enhancements} \label{section:corevsrs}

A core potential can sustain a large central density enhancement provided that $r_{\rm s} \gtrsim \varpi_{\rm c}$.
This condition can be rewritten as $GM/\varpi_{\rm c} \gtrsim \left(c_\infty^2 + u_\infty^2\right) \equiv V^{2}$, or
alternatively, as $\sigma_{\rm v} \gtrsim V$ where
$\sigma_{\rm v}$ is the velocity dispersion of the system.

In Figure \ref{fig:radii}} we examine the ability of a core potential to collect ambient gas
as a function of  the characteristic radius $r_{\rm s}$.
A density enhancement is observed to persist  as long as  $r_{\rm s} \gtrsim \varpi_{\rm c}$.
Once $r_{\rm s} <  \varpi_{\rm c}$, the collective potential is unable to alter the properties of the flow within $r\leq \varpi_{\rm c}$.

In addition, the equation of state alters  the compressional properties of the incoming  flow.  Isothermal  flows, corresponding to
$\gamma \approx  1$, allow for the effective  build up of mass.
Adiabatic gas, $\gamma = 5/3$, will instead rapidly heat the gas as it  accumulates in the core,
and halt the gas inflow at much lower
central densities than for the $\gamma\approx 1$ case.  Figure \ref{fig:mainsimple} compares the simulated density enhancements results with  those predicted by equation (\ref{corerho}).

\subsection{Bow Shock Instabilities}

Most shock instabilities identified and studied in astrophysics are related to the acceleration or deceleration of the shock itself (e.g. Bernstein \& Book 1978, Vishniac 1983).
By definition, stationary shocks are stable with respect to these mechanisms, but can still be unstable. Nakayama (1993) pointed out the radial instability of the shock if the flow is immediately accelerated after the shock transition, in isothermal flows.  In this respect the shocked Bondi flow should be unstable. The validity of the postshock acceleration criterion for adiabatic flows is still uncertain, even for radial perturbations (Nakayama 1994). Independently of the sign of post-shock acceleration,
Foglizzo \& Tagger (2000) and Foglizzo (2001) revealed another generic instability mechanism based on the cycle of acoustic and vortical perturbations in the  shock region.

The vortical-acoustic instability is fundamentally non-radial. In this case, both vorticity and entropy perturbations are advected from the shock to the accretor, and both are coupled to the acoustic perturbations. The most unstable cycle involves high frequency acoustic waves, those able to explore the hottest parts of the flow but still be refracted out, with a wavelength slightly larger than the smallest size of the sonic surface.   In the isothermal limit ($\gamma\rightarrow 1$), where entropy perturbations are absent from the problem, vorticity is more appropriate than entropy to describe the advective-acoustic cycle.

The entropic-acoustic cycle is expected to be an efficient instability mechanism for $\gamma\approx 5/3$, provided $\mu_\infty \gtrsim 2$ and that the distance to the sonic surface (with depends on $r_{\rm s}$) is small enough (Foglizzo et al. 2005). This coincides with our numerical simulations of core accretion which  show strong instability when $r_{\rm s} \approx  \varpi_{\rm c}$ (Figure~\ref{fig:radii}).  The lack of strong instability of adiabatic flows around core potentials with $r_{\rm s} \gg \varpi_{\rm c}$  could  be due to the lack of  effective acoustic feedback from the advected vorticity perturbation. Our numerical experiments  also show the presence of  instabilities within fast moving, isothermal flows. In particular isothermal flows appears much more unstable when $r_{\rm s} \approx  \varpi_{\rm c}$ (Figure~\ref{fig:gammas}).
In this conditions, a small pressure perturbation of the shock is able to generate vorticity perturbations very efficiently for $\gamma=1$ strong shocks. The differences seen in  the shock geometry between these two limiting cases precludes an accurate comparison of the stability of the flow, especially since the instability threshold is very sensitive to the size and shape of the sonic surface (Foglizzo et al. 2005).

\section{Astrophysical Relevance} \label{section:real}

The ability of core potentials to collect ambient mass depends sensitively on their velocity dispersion, relative
speed with respect to the ambient gas and gas cooling properties.  Estimates of velocity dispersion and
environmental gas properties are known for a wide range of systems,  including star clusters,
dwarf galaxies, large galactic halos, and clusters of galaxies.
In what follows, we refine our calculations to
include a wide variety of embedding media and discuss how environmental properties conspire to enable or disable the effective collection of surrounding mass
in these systems.  We assume the individual stellar members are
accreting at low enough rates such that  mass removal (and feedback) can be neglected.
These results are summarized in Table \ref{table:astros}.
\subsection{Star Clusters}
Star clusters are generally well described  by Plummer potentials \citep{plummer, bruns, pf}.  In this 
section, we will discuss two limiting cases spanning the range of environments 
star clusters might inhabit.  First, we discuss a compact star cluster moving slowly 
with respect to cold gas, representative of young  clusters in a galaxy merger environment.  This
situation is simulated in Figure \ref{fig:modelAB} (model A).  
Second, we model a less compact cluster moving quickly with respect to the 
background gas.  These less favorable conditions for gas accumulation are 
simulated in Figure \ref{fig:modelAB} (model B).

\subsubsection{Young Star Clusters}

Young massive star clusters are frequently found in merging systems like the Antennae and Cartwheel 
galaxies \citep{whitmore,cartwheel}.  In the Antennae complex, for example, young (compact)  
star clusters are observed to move slowly with respect to the  cold gas:  $c_\infty \approx  10 \, \rm{km\;s^{-1}}$ 
and $\mu_\infty \lesssim 2$ \citep{whitmore2,gandg}.  
Observations in the infrared suggest large amounts of dust present in the gas \citep{brandl}, which 
justifies  an assumption of isothermal  gas.  A typical 
cluster, $M_{\rm p} = 3.5 \times 10^{5} \, M_\odot$, $\varpi_{\rm c} = r_{\rm c} = 1 \, \rm{pc}$, in this cold, 
efficiently cooling  gas ($\gamma=1.01$), moving at $\mu_\infty = 2.0$ is shown  in Figure \ref{fig:modelAB} (model A).  In these merging environments, 
conditions are ideal for accumulating large amounts of ambient gas in these systems as $r_{\rm s} \gg r_{\rm c}$. Taking $\rho_\infty \approx 10^{-22} \, \rm{g\;cm^{-1}}$, similar to that  inferred 
by   \citet{zhu}, we  find the collected mass to be a sizable fraction of the total cluster mass: $M_{\rm acc}/M_{\rm p} \approx 0.05$.

For these young clusters (< Gyr), one might worry about the applicability of our results, as 
feedback due to massive stellar winds is neglected in our treatment.  We can estimate the magnitude of this effect by 
comparing the ram pressure 
from stellar winds of the cluster members to the pressure support generated by the collected ambient gas.
The ram pressure due to expanding stellar winds is approximately given by $P_\ast = \rho_\ast v_\ast^2 = N \dot{M}_\ast v_\ast/ (4 \pi r^2)$
for $N$ massive stars. Each star here is assumed to have a mass loss rate and wind velocity $\dot{M}_\ast$ and $v_\ast$, respectively, expanding 
into a spherical volume of radius $r$.  We can then calculate the size of the wind bubble, $r_{\rm w}$, by 
equating $P_\ast (r_{\rm w})=P_\rho(r_{\rm w})$, where $P_\rho = c_\infty^2 [\rho(r)]^{\gamma}/\gamma \approx c_\infty^2 \langle \rho \rangle$, where 
 $\langle \rho \rangle$ is the average density in the cluster's core.  Solving for $r_w$  gives $r_{\rm w} \approx 0.02 N_2^{1/2}\dot{M}_{\ast,-7}^{1/2}v_{\ast,3}^{1/2}\langle \rho \rangle_{-18}^{-1/2}c_{\infty,1}^{-1}\;{\rm pc}$, 
 where $N=10^{2}N_2$, $\dot{M}_\ast= 10^{-7}\dot{M}_{\ast,-7} \,M_\odot\;{\rm yr^{-1}}$, $v_\ast=10^3\,v_{\ast,3}\;\rm{km\;s^{-1}}$, $\langle \rho \rangle=10^{-18}\,\langle \rho \rangle_{-18}\, \rm{g\,cm^{-3}}$ and $c_\infty=10\,c_{\infty,1}\,\rm{km\;s^{-1}}$. For large density enhancements, we have  $ r_{\rm w} \ll r_{\rm c} \approx 1 \, \rm{pc}$ and, therefore  the stellar winds would be  unable to significantly  impede  the inflowing of gas.

\subsubsection{Globular Clusters}

Globular cluster's orbits are primarily dictated by the host galactic potential \citep{harris,harrisrev}.
In the Milky Way, globular clusters (GCs) are observed 
to move at highly supersonic velocities $\mu_\infty \sim 5-10$ \citep{dine}.  
In addition, the gas within the galactic halo is expected to be virialized so that $c_\infty \approx u_\infty$, implying
$V^2  \gg GM_{\rm p}/\varpi_{\rm c} $.  
As a result, for a typical GC with $M_{\rm p} = 3.5 \times 10^5 \, M_\odot$
and $\varpi_{\rm c} = r_{\rm c} = 2 \, \rm{pc}$ moving at $\mu_\infty = 4.0$, one 
expects negligible  mass accumulation as $r_{\rm s} \lesssim r_{\rm c}$. This is true even if the cluster is surrounded  by  cold, 
efficiently cooling  ambient gas ($c_\infty = 10 \, \rm{km\,s^{-1}}$ and $\gamma = 1.01$) 
as shown  in Figure \ref{fig:modelAB} (model B).   In this case, for 
 a background density similar to that of the ISM, $\rho_\infty \approx 10^{-24} \, \rm{g\,cm^{-3}}$, 
the accumulated mass is only $M_{\rm acc}/M_{\rm p} \approx 2 \times 10^{-6}$.  

A GC might collect gas more effectively 
if its orbit happened to lie in the disc plane of its host galaxy.  If the GC moving  with $\mu_\infty \leq 2$ through 
the galactic disc, we might expect the mass accumulation to more closely resemble that seen for model A in Figure \ref{fig:modelAB}.   However, these 
conditions  are rare for GCs \citep{dine}.

\subsection{Galactic Systems}

Hernquist potentials \citep{hernquist}
are used to describe a wide variety of galactic systems - from the dark 
matter halos of dwarf galaxies and Milky Way sized systems, to the halos of galaxy clusters 
\citep{dgh, gh, gch}.  In what follows we assume, for simplicity, that  these gravitational potentials are not significantly 
modified from their dark matter profiles by the presence of gas or stars.

\subsubsection{Dwarf Galaxy Halos}

Dwarf galaxies (DGs) are seen either moving through  the IGM or embedded 
within larger halos \citep{diemand, mateo}.  For those moving under the influence of a larger gravitational potential 
we expect that, on average, DGs to orbit  around the ambient gas 
with a velocity comparable to the circular velocity of their
host: $V^2 \gtrsim GM_{\rm h}/a$. However, many DGs are found to reside in nearly radial orbits, and thus will 
spend only a small  fraction of their orbital time moving at lower velocities \citep{diemand,diemand07}.  

For example, a DG  on a highly elliptical orbit, 
$e = 0.9$, will spend approximately $1\%$ of its orbital period moving at less than 
$20\%$ of the host's circular velocity.   During 
this period a large  central density enhancement is expected  to be effectively sustained  within the DG' core  if
$G M_{\rm h}/a \gtrsim c_\infty^2 + u_\infty^2$. 
That is, if the DG is surrounded by cold gas with  $c_\infty^2 \lesssim GM_{\rm h}/a$.  Such an ideal situation for mass accumulation is illustrated by 
model C in Figure \ref{fig:modelCD}.  In these conditions, if the density of cold gas is 
$\rho_\infty = 10^{-25} \, \rm{g\,cm^{-3}}$, the mass collected by  the DG 
could be as large as\footnote{We note here that model C has been run
until a quasi-static  density profile has been established, which takes about  a few core sound crossing 
times.  If the DG is at apogee for a shorter time, the accumulated mass is expected to be lower.} $M_{\rm acc}/M_{\rm h} \approx 0.14$. 
In contrast, if the surrounding gas is 
 virialized, $ c_\infty \approx \sigma_{\rm v,host} \gg G M_{\rm h}/a$, the DG halo will spend most of its orbit
with  mass inflows similar to those depicted  for model D in Figure \ref{fig:modelCD}.  In this case, for  $\rho_\infty = 10^{-25} \, \rm{g\,cm^{-3}}$,
the accumulated mass is only $M_{\rm acc}/M_{\rm h} \approx5 \times 10^{-4}$.

\subsubsection{Halos in Merging Systems}\label{hmerger}

A massive galactic halo is likely to experience a relatively equal mass merger  during its 
lifetime. Galaxies of $M \approx 10^{11} \, M_\odot$ will have undergone, on average,  about one  
merger with a galaxy of $M \gtrsim 5 \times 10^{10} \, M_\odot$ since $z=1.2$ \citep{gmergers}.  Because 
the smaller component  is moving under the gravitational influence of a halo of similar mass, we expect $u_\infty \approx  \sigma_{\rm v}$. 
Moreover, if gas is effectively virialized within  more massive halo,  then  $ c_\infty \approx \sigma_{\rm v}$ and, as a result, a large density 
build up is expected to be effectively retained  in the core of the smaller halo. This situation is depicted in Figure \ref{fig:modelE} (model E), in which a 
a Milky Way-like  sized halo is simulated moving  through hot gas with
$c_\infty = 100 \, \rm{km/s}$ at $\mu_\infty = 2.0$.   Comparing the bremsstrahlung cooling and the core
sound crossing time,  we find $t_{\rm cool} \gtrsim t_{\rm dyn}$.  Since cooling is expected to 
be inefficient, we assume  $\gamma = 5/3$.  While $r_s \gtrsim \varpi_c = a$,  ineffective  cooling 
results in an accumulated mass of only $M_{\rm acc}/M_{\rm h}= 10^{-2}$ for ambient densities of 
$\rho_\infty = 10^{-27} \, \rm{g\,cm^{-3}}$.

\subsubsection{Galaxy Clusters}

Similar to the case discussed above, a large central gas density 
enhancement is expected in equal mass merging galaxy clusters, such as the Bullet cluster complex.   
The Bullet cluster is a galaxy cluster merging with another slightly more massive  cluster \citep{mark2}.  
In this case,  
$M_{\rm h} = 10^{15} \, M_\odot$, $\varpi_{\rm c} =a= 1 \, \rm{Mpc}$, $\mu_\infty = 2.5$, and $c_\infty = 1580 \, \rm{km\,s^{-1}}$.
For an inefficiently cooled gas ($\gamma = 5/3$), this results in $r_{\rm s} \approx \varpi_{\rm c} = a$.
Such situation is depicted in Figure \ref{fig:modelF} (model F). Assuming IGM densities of $\rho_\infty = 10^{-27} \, \rm{g\,cm^{-3}}$ results in 
an accumulated mass of $M_{\rm acc}/M_{\rm h}= 3.5 \times 10^{-4}$.  This simple model of the Bullet cluster gives  similar flows structures  to those found in more
sophisticated calculations  which include the stellar and gas mass distributions.

\section{Discussion}

In \S \ref{section:real} we have shown the conditions required to collectively accumulate large 
quantities of gas from the ambient medium in the centers of a variety of astrophysical systems thought to be well
described by a core gravitational potential.  The consequences of large mass 
accumulation in such systems fall into two basic categories.  First, if the gas cools 
effectively star formation can be triggered, generating new stellar members in the system.  
Second, if the gas accretes efficiently onto the core's current stellar population, the state  of these 
stellar systems (and possibly the  central black hole) can be significantly perturbed. 
  In what follows we discuss the implications of gas accumulation 
and  their accompanying  observational signatures.

\subsection{The State of  the Accumulated Gas}

\subsubsection{Emission Properties}
The emission properties of   the accumulated gas in clusters of galaxies and young stellar clusters  are briefly reviewed in this section.
At temperatures $\gtrsim  3\times 10^7$ K,  the primary cooling process for intracluster gas is thermal bremsstrahlung, while for temperatures $\lesssim 3 \times10^7$ K, line cooling becomes very important.  For the the intracluster gas in most star clusters, the  cooling time is longer than the Hubble time (age of the universe). 
Thus cooling is not dynamically very important in these cases. Figure  \ref{fig:emiss} (bottom panel) shows the surface brightness profile, calculated assuming optically thin cooling resulting from bremsstrahlung and collisionally excited metal transitions,  
for  a cluster merging  environment  with parameters selected to
mimic the Bullet Cluster complex (model F).  We  find an integrated bolometric luminosity 
of $L_{\rm bol} \approx 10^{44} \, \rm{erg\, s^{-1}}$ and a bowshock structure that are comparable to those
observed by {\it Chandra} \citep{mark,mark2}. In the center of the cluster,  where cooling is important, our calculated emission is considerably harder than 
found in simulations including the cooling term in the energy equation \citep{milo}.

Also shown in Figure \ref{fig:emiss} is the surface brightness profile of the accumulated gas  in a young  star cluster with parameters 
similar to those seen for the  stellar cluster members in the  Antennae galaxy merging system (model A). To calculate the emission we assume optically thin cooling, where we employ an  effective cooling curve.  In the cluster's core, the near infrared  (NIR) surface brightness is 
approximately $\approx 10^{41} \, \rm{erg \, s^{-1} \, pc^{-2}}$, which is probably optimistic given that we do not explicitly  include the cooling term in the energy equation but instead use an adiabatic simulation. This simple model
gives, nonetheless, results that are  consistent with the observed range of  luminosities  in the Antennae star clusters: 
$L_{\rm NIR} = 10^{39}- 10^{42} \, \rm{erg \, s^{-1}}$ \citep{brandl}.  For clusters embedded in these types of merging  environments, the 
emission from the bowshock and cluster core should distinguishable\footnote{This is because the change in brightness across these two components is at least two orders of magnitude.} given enough spatial resolution and provided the   reprocessed  stellar radiation is not the dominant contribution to the NIR. 

\subsubsection{Enhanced Central Densities Leading to Star Formation}
If the gas collected by these potentials cools efficiently, we would expect 
it to fragment and form stars if the density is high enough.  In general, one can relate 
the gas density to the star formation rate \citep{ks}: 
$\Sigma_{\rm sfr} \approx 10^{-4} $ $(\Sigma_{\rm gas}/10^4 \, M_\odot {\rm \,pc^{-2}})^{1.4} $ $
M_\odot \, {\rm yr^{-1} \, pc^{-2}}$,
where $\Sigma_{\rm sfr}$ and $\Sigma_{\rm gas}$ are the star formation and gas column
 densities, respectively. A dwarf galaxy, for example, embedded within a larger halo with properties similar to those 
 assumed for model C  (i.e., $\rho_\infty = 10^{-24} \, \rm{g \, cm^{-3}}$ and a $200 \, \rm{pc}$ star forming region) is expected to  form  stars at a rate of $\Sigma_{\rm sfr} = 10^{-1} \, \rm{M_\odot \, yr^{-1}}$,  similar to that predicted 
by star formation simulations in these systems \citep{sfrg}.  

For the star clusters embedded in the  Antennae galaxy merging system, whose conditions are relatively well described by  our model A, 
we find  $\Sigma_{\rm sfr} \approx  4 \times 10^{-2} \, \rm{M_\odot \, yr^{-1}}$ per system, or 
a total rate of $\Sigma_{\rm sfr} \approx 4 (N_{\rm c}/10^2)\, \rm{M_\odot \, yr^{-1}}$ for  all $N_{\rm c}$  clusters. Such star formation rates were calculated assuming a star forming region of  $5 \, \rm{pc}$ and a background ambient  density of $\rho_\infty = 10^{-21} \, \rm{g \, cm^{-3}}$, which results in an average increase in cluster 
gas core density of $\langle \rho \rangle = 10^{-18} \, \rm{g \, cm^{-3}}$ (model A). The resulting  rates are similar to those observed in the Antennae  complex \citep{gandg}.  If we assume that such rates are effectively  sustained over the calculated  orbital passage time scale of  about 20~Myrs \citep{naab},  one expects at least  $10^{5} \, \rm{M_\odot}$ of new stars to be  created per system. 
More detailed  simulations that take into account self-gravity,  realistic cooling and feedback effects are needed before detailed comparisons can be made with the inferred conditions  required to  explain  subsequent 
episodes of star formation in stellar clusters \citep{conroy, gratton, carr, piotto, mack}.

\subsection{Accretion Onto Individual Stellar Members and the  Central Massive Blackhole}
A compact star (or a black hole) of mass $M_\ast$, moving with relative velocity $u$
through a gas of ambient density $\rho_\infty$ and sound speed $c_\infty$,
nominally accretes at the Bondi-Hoyle-Lyttleton rate: $\dot{M}\eqsim
4\pi (GM_\ast)^2\rho_\infty (u^2+c_\infty^2)^{-3/2}$. If, however, the collective potential of the system (e.g. star cluster or galaxy) was able to
significantly increase the surrounding gas density prior to being
accreted onto the individual neutron star members, then  the accretion rates of
the individual core stellar  members (or the central black hole) would be enhanced by $\sim \langle \rho \rangle/\rho_\infty$, where $\langle \rho \rangle$ is the average increase in gas density in the core. It is to this problem that we
now turn our attention. In particular,  we study the effects of enhanced accretion rates 
onto the  white dwarf core members  and the  central massive black hole.

\subsubsection{White Dwarfs in Stars Clusters and  Galaxies}

In mature stellar systems, we expect a significant fraction of the stellar mass to be in
white dwarfs.   If the collective potential of the system (which contains the white dwarfs) is able to alter the ambient  gas properties before the gas 
is accreted onto the individual stars,  then the resulting enhanced accretion rates could significantly alter the state of the accreting stellar population.
If, however,   the  density   is not significantly
increased, then the stars accrete gas as though they move through the
external medium independently. 

The resulting mass accretion rates for  the white dwarf members  depends
on the radial distribution of both stars and gas. Here we calculate the expected mass accretion rates
in the core of  slowly moving star cluster (model A)  using two extreme examples for the radial
distribution of white dwarfs remnants. The first one is based on
Fokker-Planck models of a core collapsed (centrally condensed)
globular cluster \citep{dull}, 
and the second simply
assumes that the white dwarfs, containing 1\% of the total
mass, follow the radial stellar mass distribution. 
Figure \ref{fig:wdmdot} shows the mass accretion rate distribution of the
white dwarf cluster members for both radial distributions (solid and dashed lines respectively).

This range of mass accretion rates  are expected to result in enhanced novae rates \citep{nomoto}.
With a nova outburst rate of $\Psi_{\rm nr}\approx 10^{-4}\;{\rm yr^{-1}}$, we expect
about $\approx 0.1 (\Psi_{\rm nr}/10^{-4}\;{\rm yr^{-1}})$ $(N_{\rm wd}/10^4)(N_{\rm c}/10^2)$ outbursts per year.
This is probably a conservative estimate as  the individual novae rates are  expected to increase  to $\Psi \leq  10^{-1}\;{\rm yr^{-1}}$ 
for the larger accretion rates calculated here \citep{starrfield}.  These novae outbursts will be observed  as super 
soft X-ray sources ($kT\le 1\rm{keV}$) with luminosities $\lesssim 10^{38} \, \rm{erg\, s^{-1}}$ and 
thus might be observable in nearby merging galaxies \citep{hernanz}.  

Such high mass accretion rates can  only be sustained if star clusters  move 
slowly with respect to the background gas.  
While there are no stellar clusters observed in the Galactic disk
which bear these anticipated properties (the relative velocity of the
halo clusters to the interstellar medium is in the range of 100 km s$^{-1}$),
observations of several cluster knots in the Antennae indicate
intracluster relative velocities that comparable to the central
velocity dispersions.  In these systems,   it has been shown 
that this merging state is, however,  short lived, $\Delta t \lesssim 20 \, \rm{Myr}$ \citep{naab}. Even if
this (slowly moving) cold gas phase persists for  $\leq 20 \, \rm{Myr}$, we expect  
at least 30\% of the white dwarf members to double their mass.  This increase in mass would in turn 
lead to a sizable increase in luminosity \citep[e.g.][]{mestel}, which  in turn modify  the white dwarf cooling sequence. 

\subsubsection{Accretion onto Central Massive  Black Holes}
Most  early-type galaxies and spiral bulges are now thought
to contain massive black holes in their nuclei \citep{mag}. 
Direct evidence is also now available which supports the idea
that active galaxies are powered by massive black holes. 
It can then be argued that many if not
all galaxies could be expected to have undergone an active phase
and to possess a central massive black hole \citep{handr},
although the  presence of intermediate mass 
black holes in dwarf galaxies  (and some globular clusters) remains controversial 
\citep{greene, geb, baum1}.

Large density enhancements in the nuclei of galaxies, dwarf galaxies and star clusters
would  be accompanied by enhanced mass supply to the central black hole.
For a typical  galactic halo (model E), this enhanced mass accretion results
in a significant  black hole  accretion luminosity: 
\begin{equation}
L_{\rm bh} \approx 10^{41} {M_{\rm bh} \over 10^6\,M_\odot} {\epsilon \over 0.1} \, \rm{erg \, s^{-1}},
\end{equation}
where $\epsilon $ is the efficiency for converting gravitational
energy into  radiation.  Extrapolating the $M_{\rm bh} - \sigma$ relation \citep{tremaine} to the star clusters 
and dwarf galaxies modeled by simulations [A, B, C, D] gives black hole masses of $M_{\rm bh} = [0.68, \, 0.68, \, 1.42, \, 1.42]\times10^4 \, M_\odot$ and 
accretion luminosities of $\log{L_x/L_{\rm edd}} = [10^{1.77}, \, 10^{-1.46}, \, 10^{2.18}, \, 10^{-6.20}]$, respectively.
While the integrated output from this increased mass feeding  could, in principle,
amount to large factors of the black hole's Eddington
luminosity, it would probably be significantly less because we have, for simplicity, neglected feedback. \\

We thank Thorsten Naab, Elena D'Onghia, Charlie Conroy, Aaron Romanowsky, J\"{u}rg Diemand, and Lars Hernquist  for useful discussions. We also thank the referee for his/her insightful comments. 
The software used in this work was in part developed by the DOE-supported ASCI/Alliance Center for Astrophysical Thermonuclear Flashes at the University of Chicago. Computations were performed on the Pleaides UCSC computer cluster. This work is supported by NSF: AST-0847563 (J.N. and E.R.), NASA: NNX08AL41G (J.N. and D.L.), and The David and Lucile Packard Foundation (E.R.).

\clearpage

\begin{deluxetable*}{ccccccccccl} 
\tabletypesize{\tiny}
\tablecolumns{11} 
\tablewidth{0pc} 
\tablecaption{Astrophysically Motivated Simulations} 
\tablehead{ 
\colhead{Simulation} & \colhead{$\gamma$} & \colhead{$c_\infty$} & \colhead{$\mu_\infty$} & \colhead{${\rm \varpi_c}$} & \colhead{$M_p$ or $M_h$}  
& \colhead{Analytic} & \colhead{Simulated}  & \colhead{Analytic} & \colhead{Simulated} & 
\colhead{$u_{\rm max}/c_\infty$}  \\
\colhead{Label} &       \colhead{ $\,$}                   & \colhead{${\rm [ km \, s^{-1} ]}$}  & \colhead{$\,$}   & \colhead{${\rm [pc]}$} & \colhead{${\rm [M_\odot]}$} & 
\colhead{$\log(\rho_{\rm max}/\rho_\infty)$} & \colhead{$\log(\rho_{\rm max}/\rho_\infty)$}  & \colhead{$r_{\rm s}$ ${\rm [pc]}$} & \colhead{$r_{\rm s}$ ${\rm [pc]}$} & \colhead{$\,$} }
\startdata 
 A\tablenotemark{a} &	1.01 &	10	& 2	& 1	& $3.5 \times 10^{5}$   &           		6.03 	& 4.32 	& 2.7	&	1.7    & 	3.8  \\
B\tablenotemark{b}  &	1.01	 &   10 & 	4	& 2	& $3.5 \times 10^{5}$  &	          		3.12	 & 1.08 &	1.34	  &	0\tablenotemark{$\dagger$}	&	5.3  \\
C\tablenotemark{c} &	1.01	 &    10	& 2	& 200	& $10^{8}$	&  	8.38 &	 4.79	& 1500	&	210	&	4.8 \\
D\tablenotemark{d} &	1.67	 &   100 &	4	& 200	& $10^{8} $	&  	0.09  &	0.03	 & 11	&	0\tablenotemark{$\dagger$}	&	4  \\
E\tablenotemark{e} &	1.67 &  100	& 2	& $3 \times 10^4$ &	$8 \times 10^{11}$	&  	1.5 &	0.77 &	$1.5 \times 10^5$	&	$1.4 \times 10^4$ &	
	3.23  \\
F\tablenotemark{f} &	1.67	 &   1580	& 2.5	& $10^6$ &	   $10^{15}$ &  		0.5	& 0.86	& $6 \times 10^5$ &		$3.6 \times 10^5$ &		4.4 
\enddata 
\tablenotetext{a}{Young star cluster modeled here by a Plummer potential.}
\tablenotetext{b}{Globular cluster modeled by a Plummer potential.}
\tablenotetext{c}{Slowly moving dwarf galaxy in cool gas modeled by a Hernquist potential.}
\tablenotetext{d}{Slowly moving dwarf galaxy in hot gas modeled by a Hernquist potential.}
\tablenotetext{e}{Milky Way sized halo moving through hot gas modeled by a Hernquist potential.}
\tablenotetext{f}{Simulation of galaxy cluster similar to the Bullet Cluster, modeled by a Hernquist potential.}
\tablenotetext{$\dagger$}{These simulations resulted in  a detached tailshock.}
\label{table:astros}
\end{deluxetable*}

\begin{figure}
\centering\includegraphics[width=0.5\textwidth]{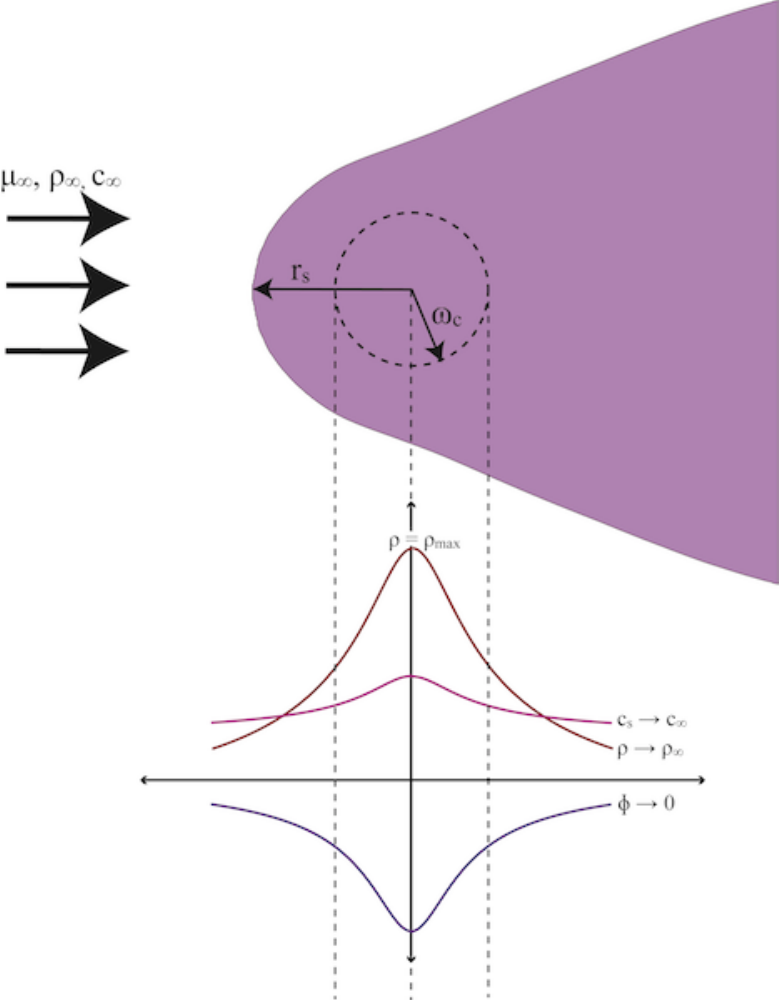}
\caption{Diagram shows the flow geometry of non-stationary core potentials with 
central mass accumulation. A bow shock
is produced by the motion of the potential with respect to the ambient gas. The initial asymptotic Mach number, sound speed and density of the inflowing gas are denoted here as $\mu_\infty$, $c_\infty$ and  $\rho_\infty$, respectively. Also depicted are the characteristic, $r_{\rm s}$, and  the core, $\varpi_{\rm c}$, radii, together with a sketch of the expected profiles of density, $\rho(r)$, and sound speed, $c_{\rm s}(r)$, along the flow axis.}
\label{fig:diagram}
\end{figure} 

\begin{figure}
\centering\includegraphics[width=0.4\textwidth]{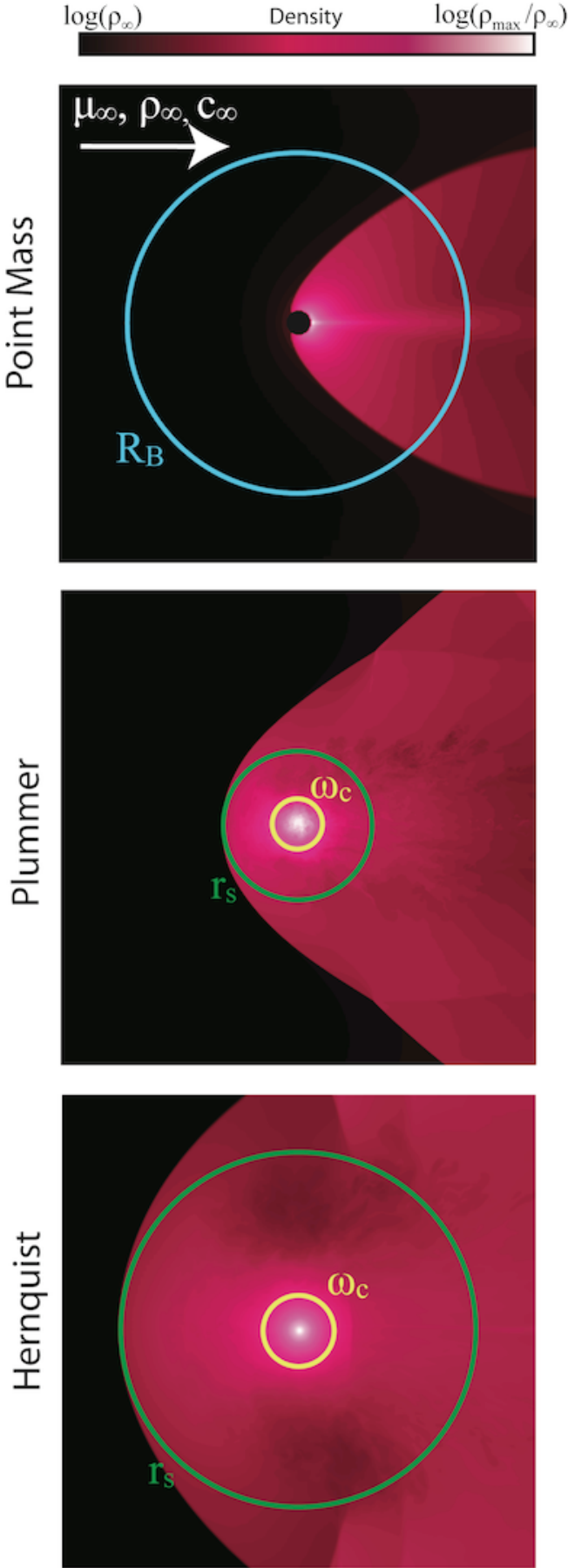}
\caption{Density contours of the flow pattern around   point mass, Plummer and Hernquist potentials set in motion through
  an initially uniform medium with $\rho_\infty=1$ and $\mu_\infty=2$.  The point mass potential is implemented here  with a sink 
boundary following the method of \citet{ruffert} by flooring the values of density and 
pressure to 50\% of their background values and setting the velocities to  zero inside the sink radius.  
Color bars show density cuts through the {\it{xy}}-plane with limiting values: $\rm {log}\; (\rho_{\rm max}/\;\rho_\infty)=[1.2,1.1,1.5]$ for point mass, Plummer and Hernquist potentials, respectively. 
The equation of state used here  is adiabatic ($\gamma=5/3$).  Also depicted are the characteristic, $r_{\rm s}$ (which reduces to the Bondi radius $R_{\rm B}$ for a point mass potential), and  the core, $\varpi_{\rm c}$, radii. The Plummer and Hernquist potentials have been constructed to yield the same circular velocity peak radius: $r_{\rm v}$. Note that  the difference  between $r_{\rm s}$ in  the  Plummer and  Hernquist potentials is consistent  with equation  (\ref{sizecomp}). }
\label{fig:comp}
\end{figure} 
\begin{figure}
\centering\includegraphics[width=0.25\textwidth]{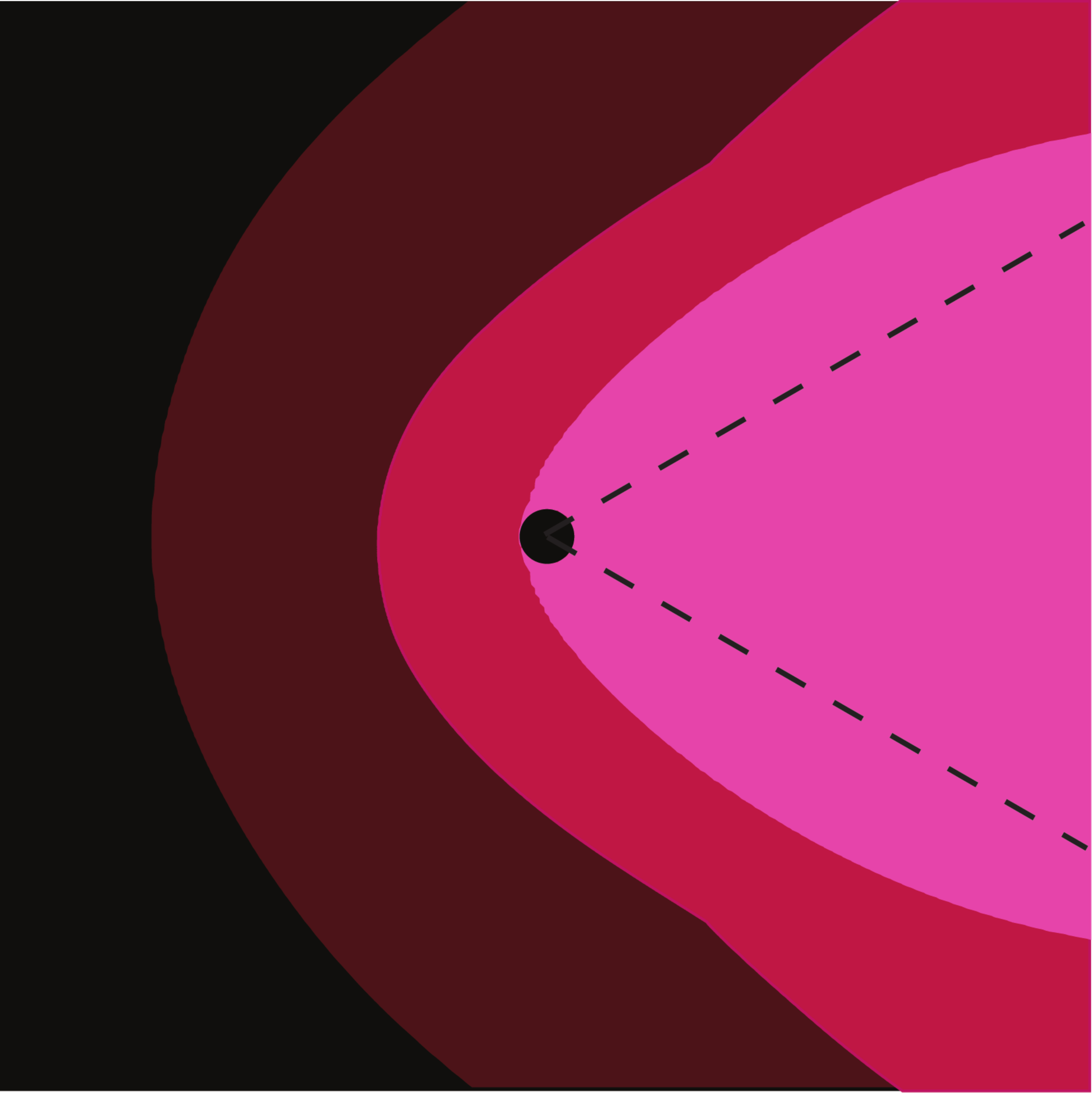}
\caption{Bow shock patterns  produced  around a  point mass (pink), Plummer  (red) and Hernquist (brown) potentials set in motion through an initially uniform medium with $\mu_\infty = 2.0$. Also plotted, for comparison,  is  the bow shock  expected for a supersonically moving, non-gravitating object (dashed line). In this case the bow shock's opening angle is simply  given by $\theta_{\rm bow} = \arcsin \mu_\infty^{-1}$. The 
size of the box is 40 times the sink size (used here to model the point mass potential), depicted by the black circle.}
\label{fig:bows}
\end{figure}

\begin{figure*}
\centering\includegraphics[width=0.8\textwidth]{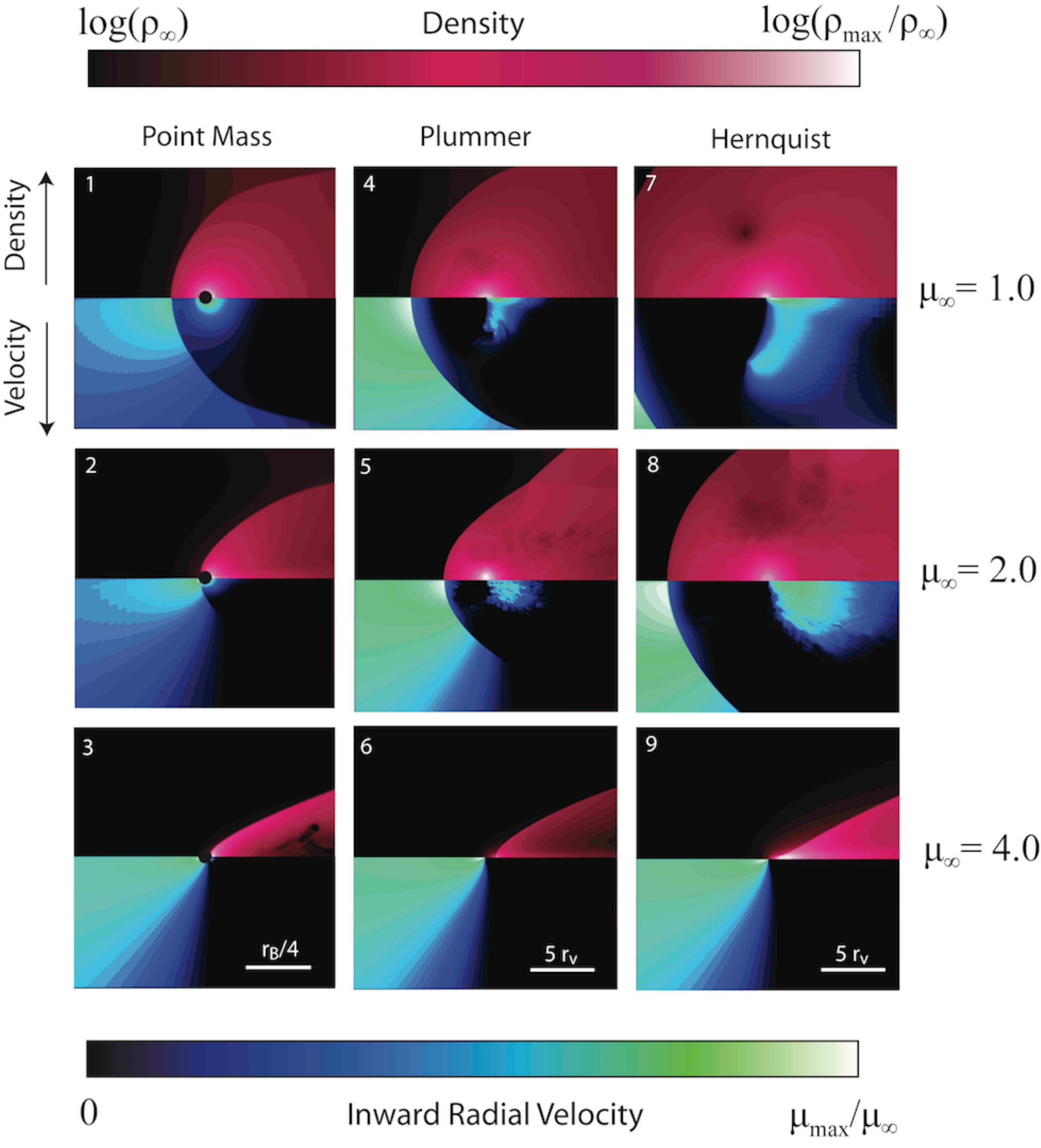}
\caption{Density and radial velocity contours for gas flowing around a point mass, Plummer and Hernquist potential at various relative velocities. We exploit the flow symmetries to plot both the density contrast (upper, maroon half) and inward radial velocity 
profiles (lower, blue half) in the same panel.
 Color bars show density and velocity cuts through the {\it{xy}}-plane in units of $\log
    (\rho_{\rm max}/\rho_\infty$) and $\mu/\mu_\infty$.  The maximum densities and radial velocities  attained in simulations 1 through 9 are  
    $\log (\rho_{\rm max}/\rho_\infty) =[1.5,1.2,0.8,1.4,1.1,1.4,1.7,1.5,1.1]$ and  $(\mu_{\rm max}/\mu_\infty) =[3.9,6.2,6.9,1.9,3.1,6.2,1.7,2.9,6.4]$, respectively.
The equation of state used here is adiabatic ($\gamma=5/3$). 
}
\label{fig:machc}
\end{figure*} 

\begin{figure}
\centering\includegraphics[width=0.5\textwidth]{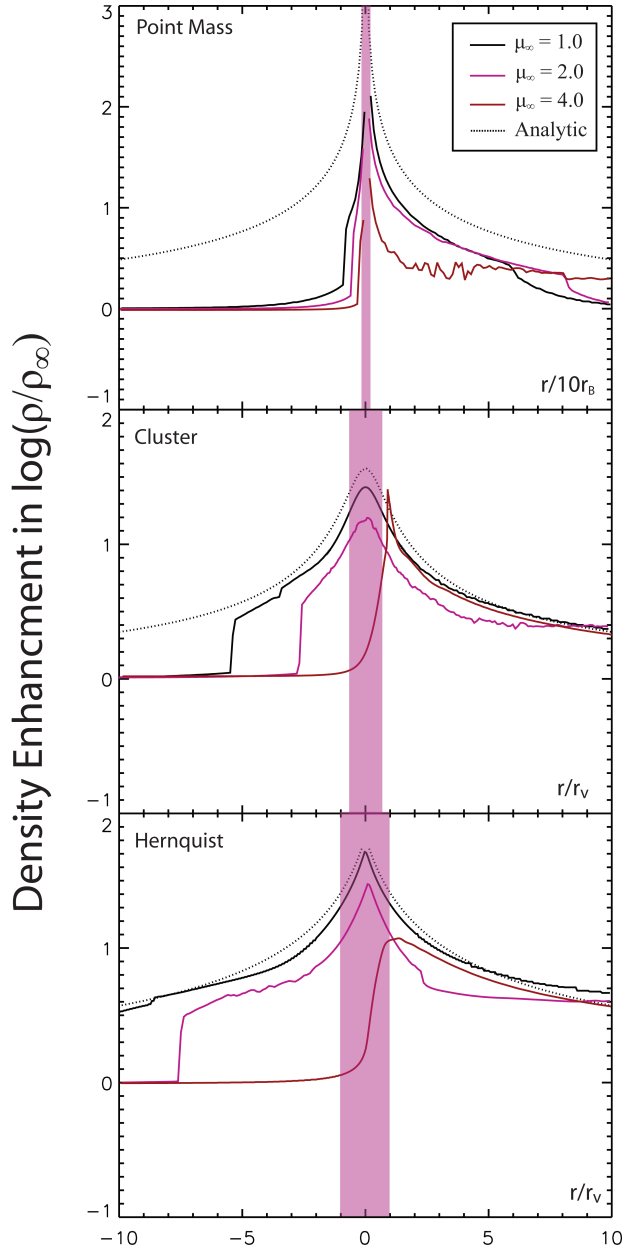}
\caption{Density cuts along the axis parallel to the incoming flow for the simulations shown in Figure \ref{fig:machc}. 
The dotted lines give the analytic solutions  for  a point mass, Plummer and Hernquist  potential,  while the black, pink and red lines show the numerical profiles for
  $\mu_\infty = 1.0$, $2.0$, and $4.0$, respectively. The shaded regions depict either the core radii or the sink size.}
\label{fig:wanal}
\end{figure}

\begin{figure}
\centering\includegraphics[width=0.5\textwidth]{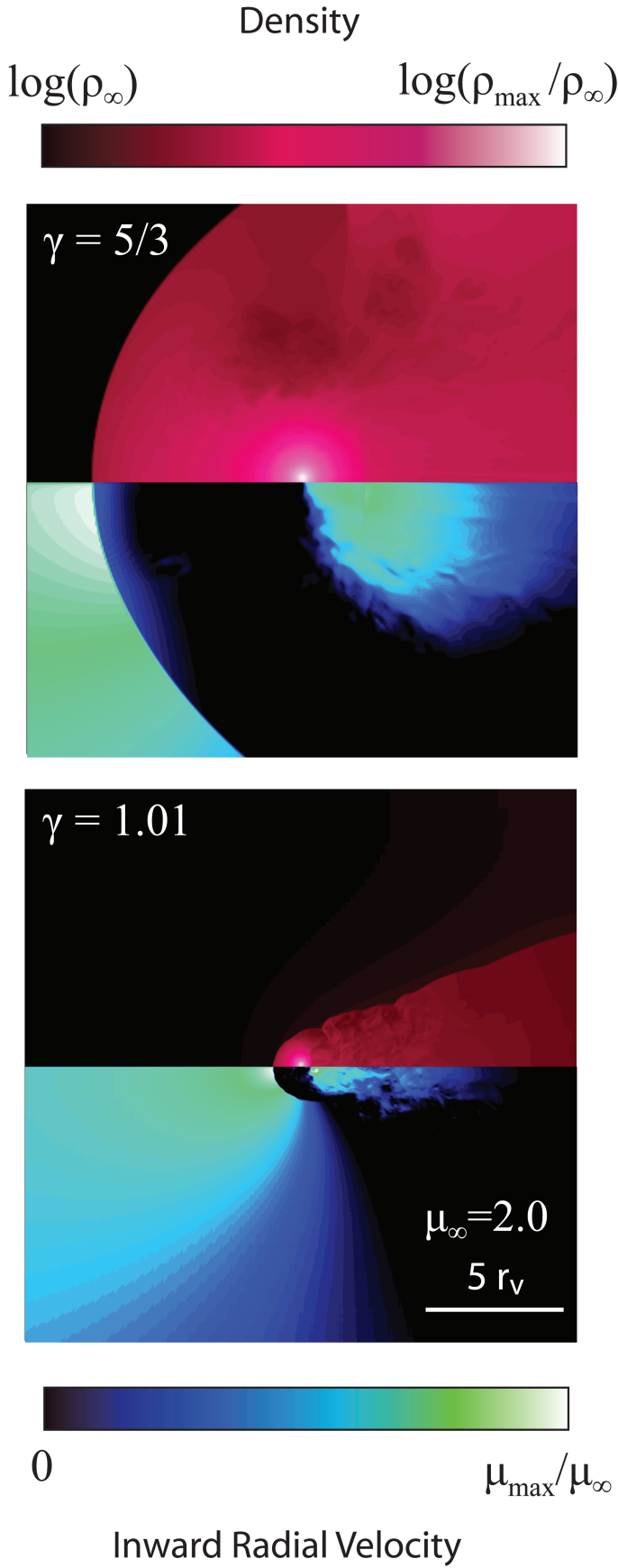}
\caption{Density and radial velocity contours for gas flowing with $\mu_\infty = 2.0$ around a   Hernquist potential 
in the adiabatic  ($\gamma = 5/3$) and isothermal ($\gamma = 1.01$) limits. The maximum densities and radial velocities  attained in simulations for $\gamma = 5/3$ and $\gamma=1.01$ are  
    $\log (\rho_{\rm max}/\rho_\infty) =[1.5,5.1]$ and  $(\mu_{\rm max}/\mu_\infty) =[ 2.9,4.9]$, respectively.}
\label{fig:gammas}
\end{figure} 

\begin{figure}
\centering\includegraphics[width=0.5\textwidth]{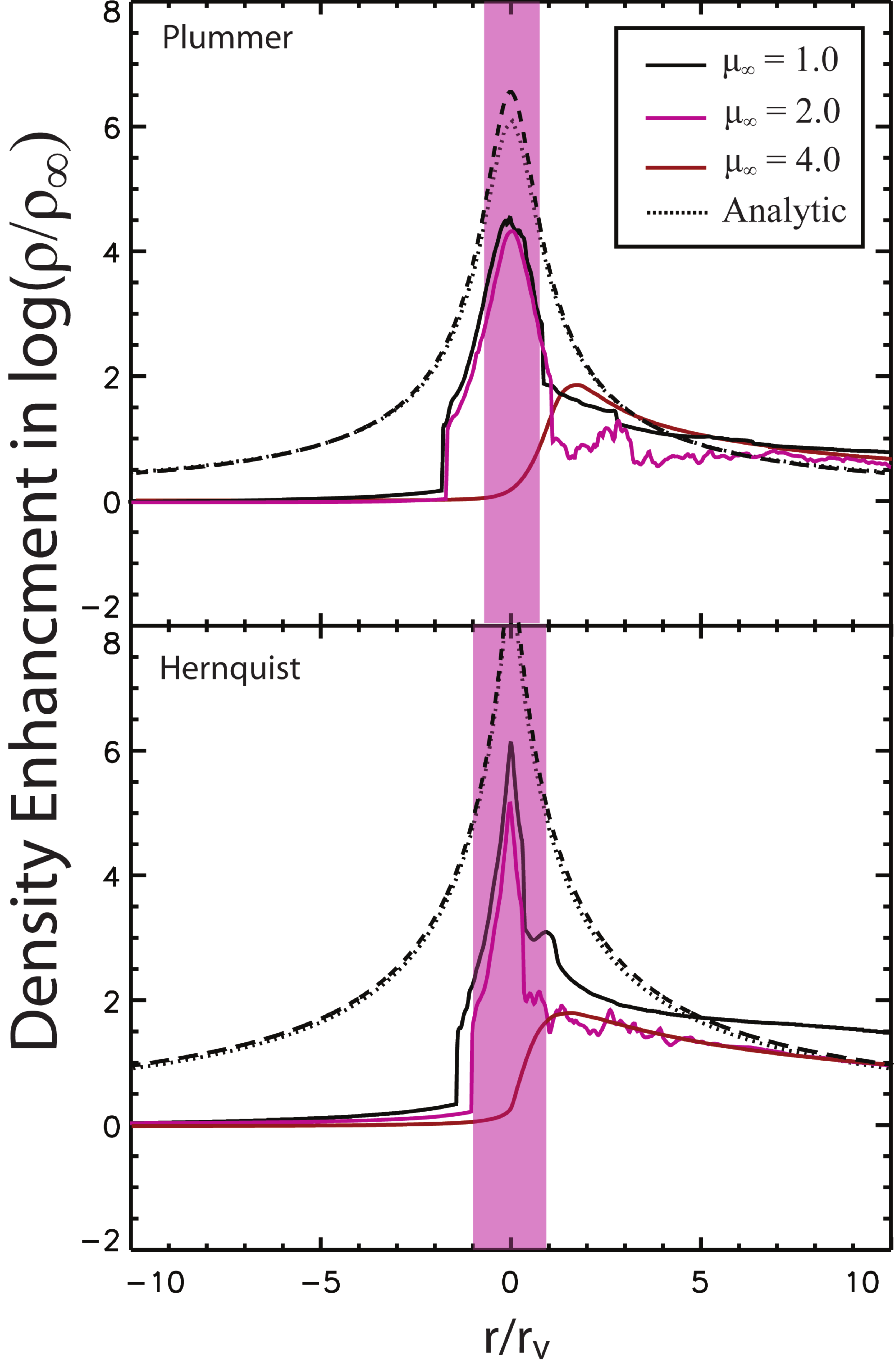}
\caption{Same as  Figure \ref{fig:wanal} but for a near isothermal equation of state ($\gamma = 1.01$).  
The dotted lines are the analytic profiles derived in \S \ref{section:analytic}, specifically  equations (\ref{eq:dcl}) and (\ref{eq:dhe}) 
 with $\gamma = 1.01$.  The dashed lines are the isothermal form of the density equation, (\ref{isodens}). 
The shaded regions show the size of the core.}
\label{fig:gammasb}
\end{figure} 

\begin{figure}
\centering\includegraphics[width=0.3\textwidth]{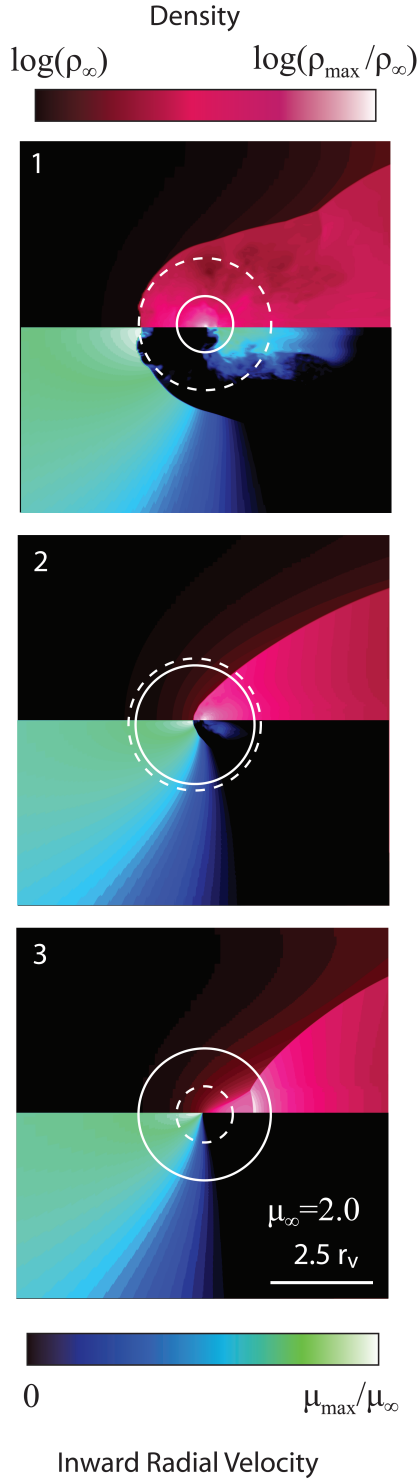}
\caption{Density and radial velocity contours for gas flowing with $\mu_\infty = 2.0$ around a Hernquist potential with varying $r_{\rm s}$
in the adiabatic  ($\gamma = 5/3$)  limit. Dashed circles depict the location of the analytic characteristic radius, $r_{\rm s}$ (equations [\ref{zeta}] and  [\ref{rs}]), while solid circles give the approximate location of the  core radius $\varpi_{\rm c}$. The maximum densities and radial velocities  attained in simulations 1 through 6 are  $\log (\rho_{\rm max}/\rho_\infty) =[1.8,1.3,1.0]$ and  $(\mu_{\rm max}/\mu_\infty) =[3.3,3.6,3.2]$, respectively.}
\label{fig:radii}
\end{figure} 

\begin{figure}
\centering
\includegraphics[width=0.5\textwidth]{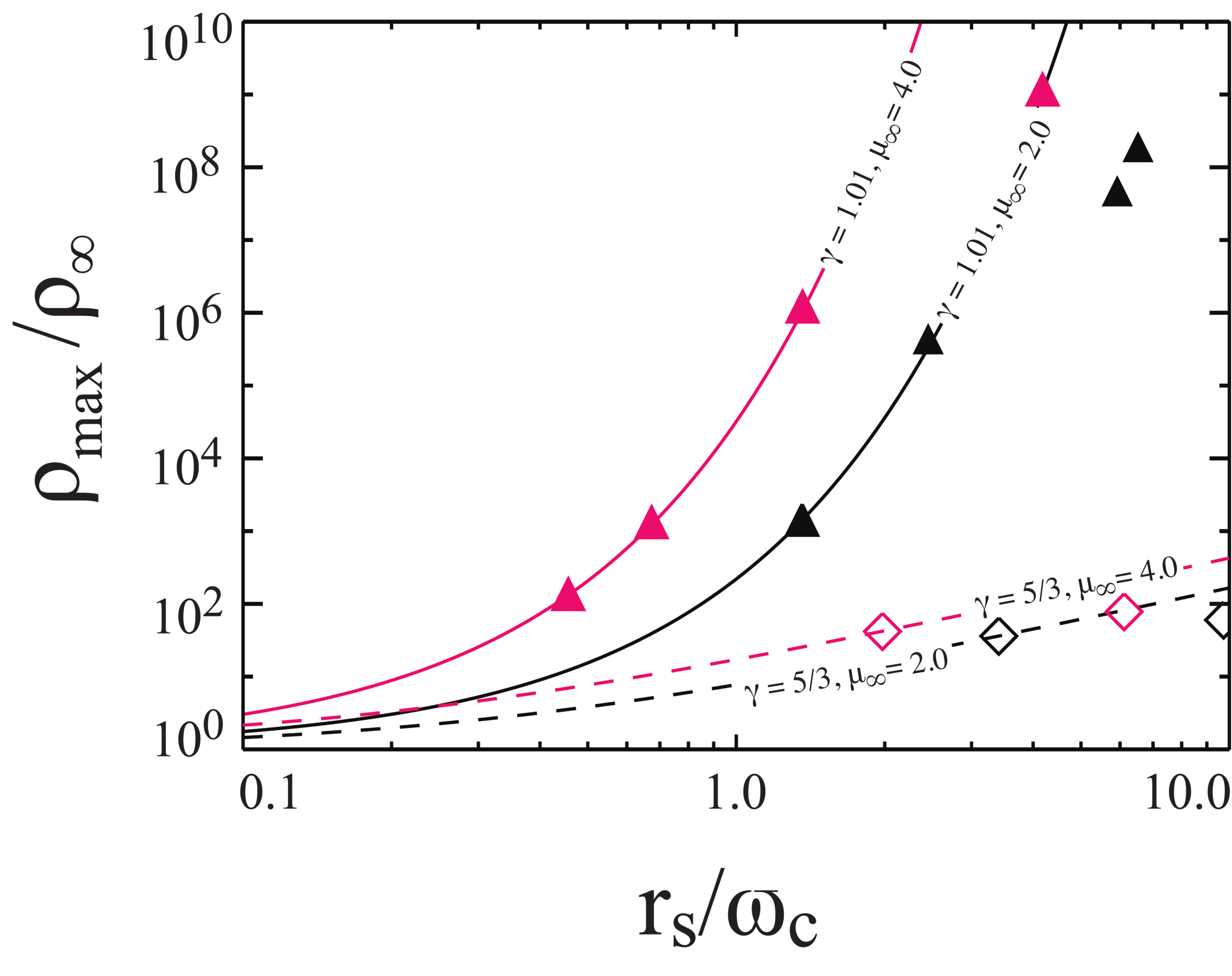}
\caption{Expected density enhancements as a function of $r_{\rm s}/\varpi_{\rm c}$ for a variety of 
simulations with $c_\infty = 10 \, {\rm km \, s^{-1}}$.  The triangles (diamonds) are measured $\rho_{\rm max}/\rho_{\infty}$ 
and $r_{\rm s}/\varpi_{\rm c}$ from the isothermal (adiabatic) simulations.  
For comparison, the lines are trends of $\rho_{\rm max}/\rho_{\infty}$ from the analytic results from equations 
(\ref{rs}) and (\ref{eq:dcl}) 
for Plummer potentials with a variety of masses and core radii in $c_\infty = 10 \, {\rm km \, s^{-1}}$ gas.}
\label{fig:mainsimple}
\end{figure}

\begin{figure}
\centering
\includegraphics[width=0.4\textwidth]{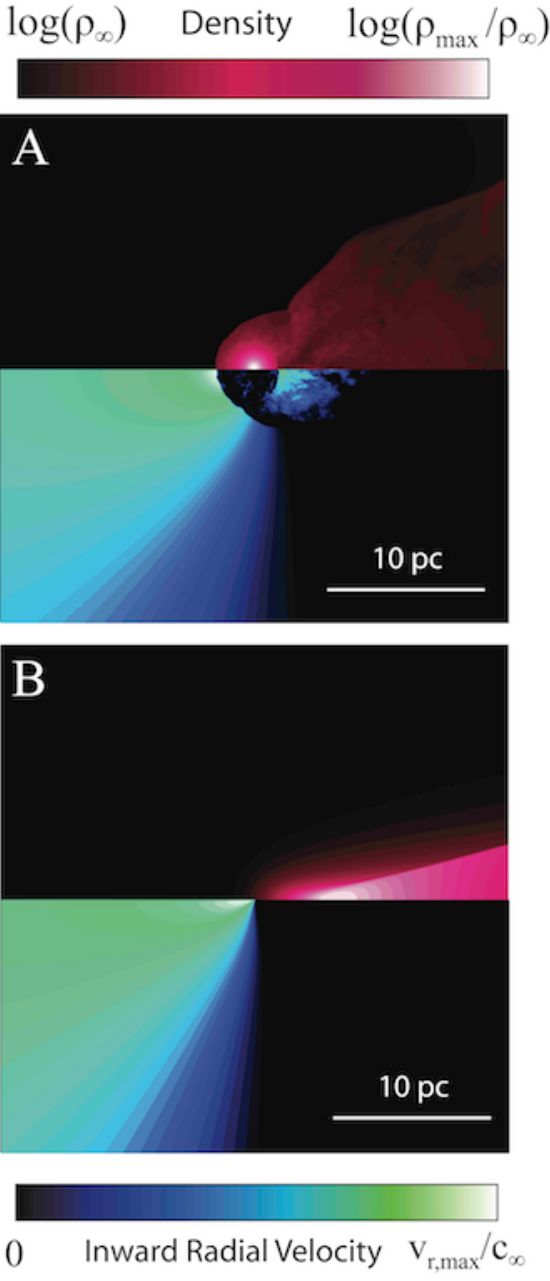}
\caption{Examples of gas accumulation in  star  clusters. 
 Density and radial velocity contours for models A and B.  
The maximum densities and radial velocities attained in the simulations  are 
$\log(\rho_{\rm max}/\rho_\infty) = [4.32, 1.08]$ and $(u_{\rm max}/c_\infty) = [3.8, 5.3]$, respectively.  }
\label{fig:modelAB}
\end{figure}

\begin{figure}
\centering\includegraphics[width=0.4\textwidth]{fig8_t2.pdf}
\caption{Examples of gas accumulation in  dwarf galaxies.   Density and radial velocity contours for models C and D.  
The maximum densities and radial velocities attained in the simulations  are 
$\log(\rho_{\rm max}/\rho_\infty) = [4.79, 0.03]$ and $(u_{\rm max}/c_\infty) = [4.8, 4.0]$, 
respectively.}
\label{fig:modelCD}
\end{figure} 

\begin{figure}
\centering\includegraphics[width=0.5\textwidth]{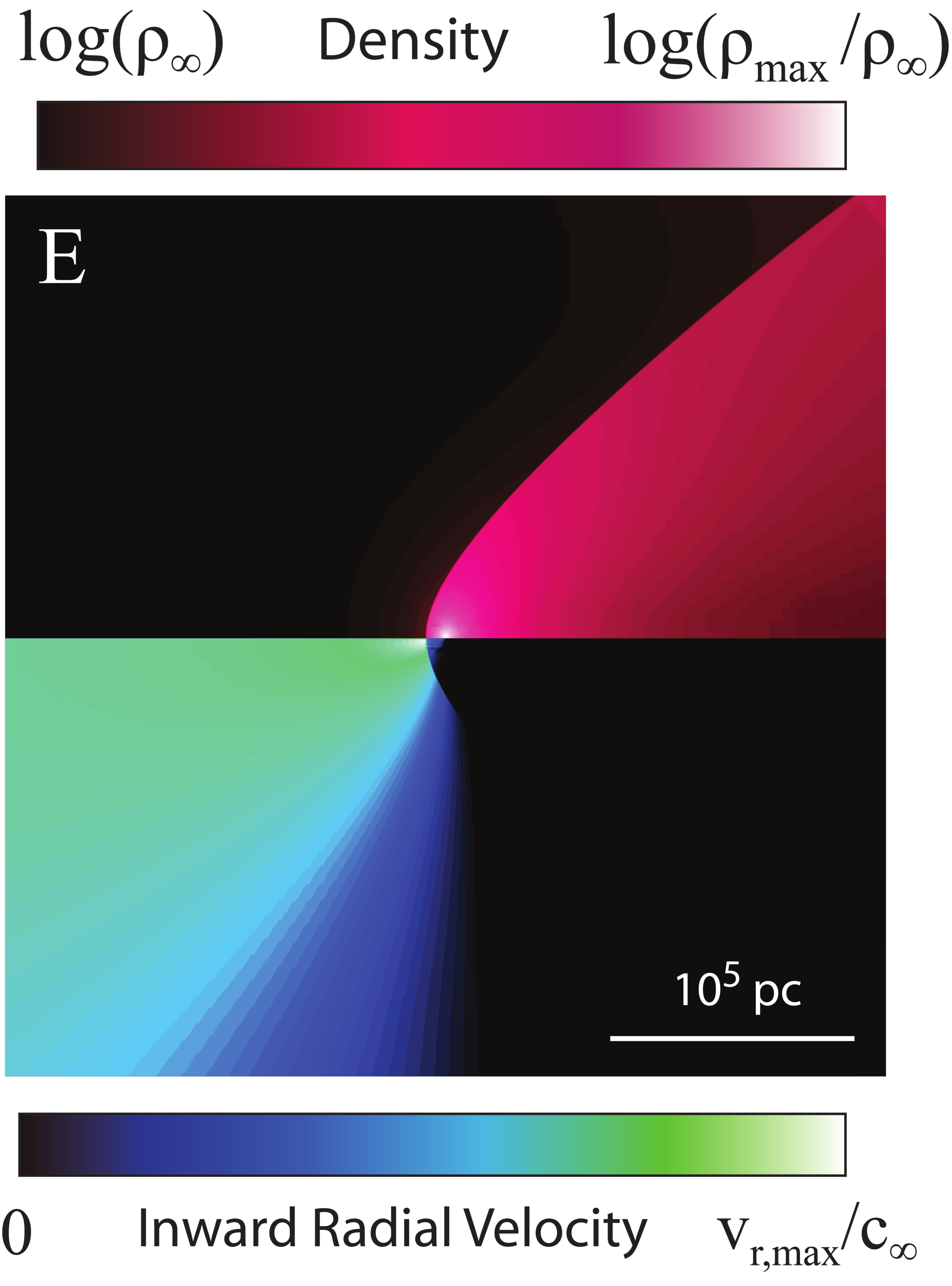}
\caption{Gas accumulation for a Milky Way-like  halo moving in a major merging environment. 
Density and radial velocity contours for model E.  
The maximum densities and radial velocities attained in the simulation are 
$\log(\rho_{\rm max}/\rho_\infty) = 0.77$ and $(u_{\rm max}/c_\infty) = 3.23$, respectively.  }
\label{fig:modelE}
\end{figure}

\begin{figure}
\centering\includegraphics[width=0.5\textwidth]{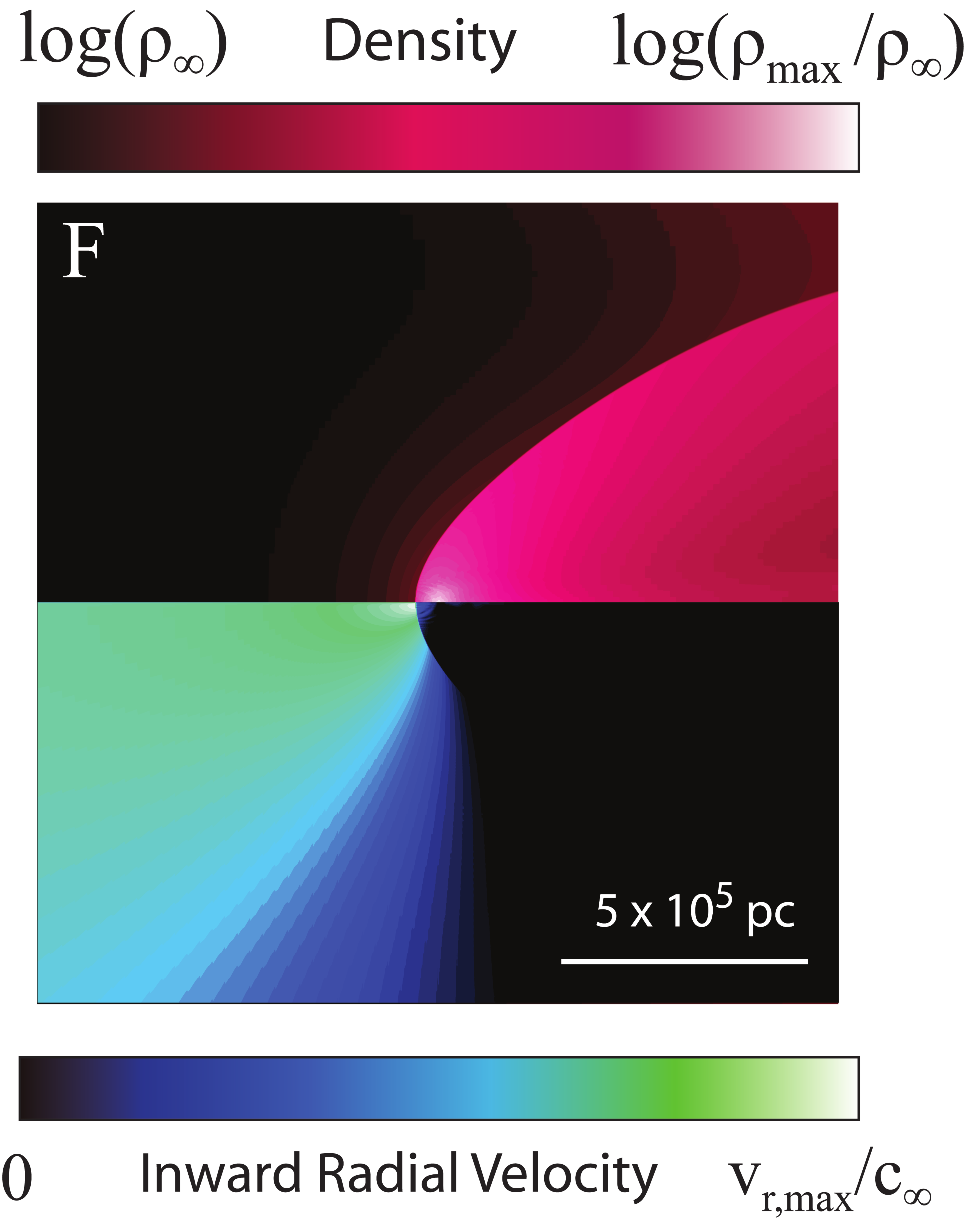}
\caption{Gas accumulation for a galaxy cluster in a major merger environment. 
Density and radial velocity contours for model F.  
The maximum densities and radial velocities attained in the simulation are 
$\log(\rho_{\rm max}/\rho_\infty) = [0.86]$ and $(u_{\rm max}/c_\infty) = [4.4]$, 
respectively. }
\label{fig:modelF}
\end{figure}

\begin{figure}
\centering
\includegraphics[width=0.5\textwidth]{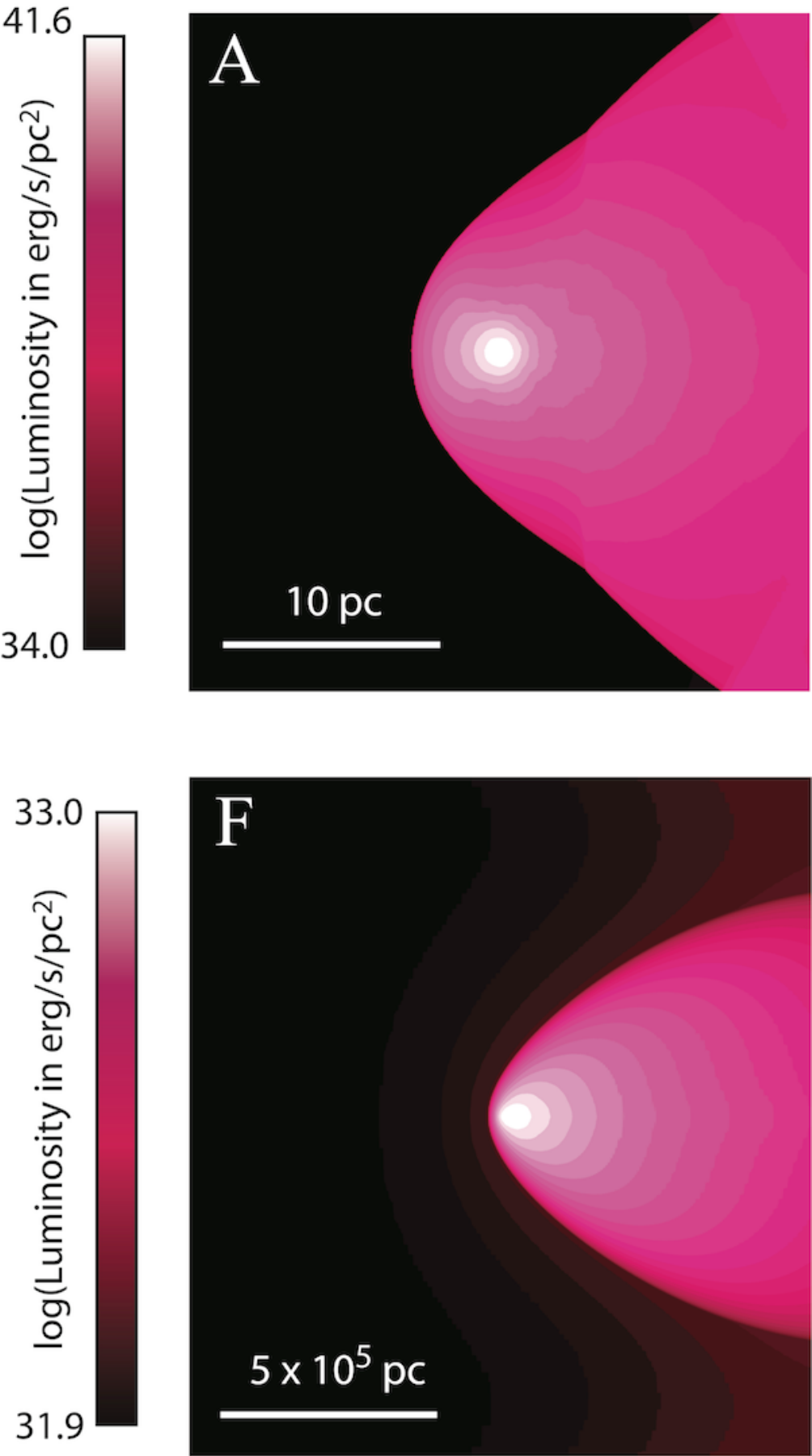}
\caption{The emission properties of   the accumulated gas in a galaxy cluster (model F)   and a young stellar cluster (model A).
Emission is calculated using the cooling curves for an optically thin plasma from \cite{peres} for chromospheric energy losses 
in the temperature range $4.44 \times 10^3 < T < 2 \times 10^4 \, K$, the curves of 
\cite{rosner} for coronal temperatures of $2 \times 10^4 < T < 10^8 \, K$, and bremsstrahlung emission above 
$10^8 \, K$. We assume $\gamma=5/3$. That is,  the cooling term is not explicitly  included in the energy equation. 
The ionization 
fraction is assumed to be purely collisional, giving $\chi_i = 0.003$ for model A 
and $\chi_i \sim 1$ for model F.  Model A
has a temperature range in the band $kT = 0.6 - 7.2 \, \rm{eV}$, while for model F
the range is $kT = 15.0 - 146.0 \, \rm{keV}$.}
\label{fig:emiss}
\end{figure}

\begin{figure}
\centering
\includegraphics[width=0.5\textwidth]{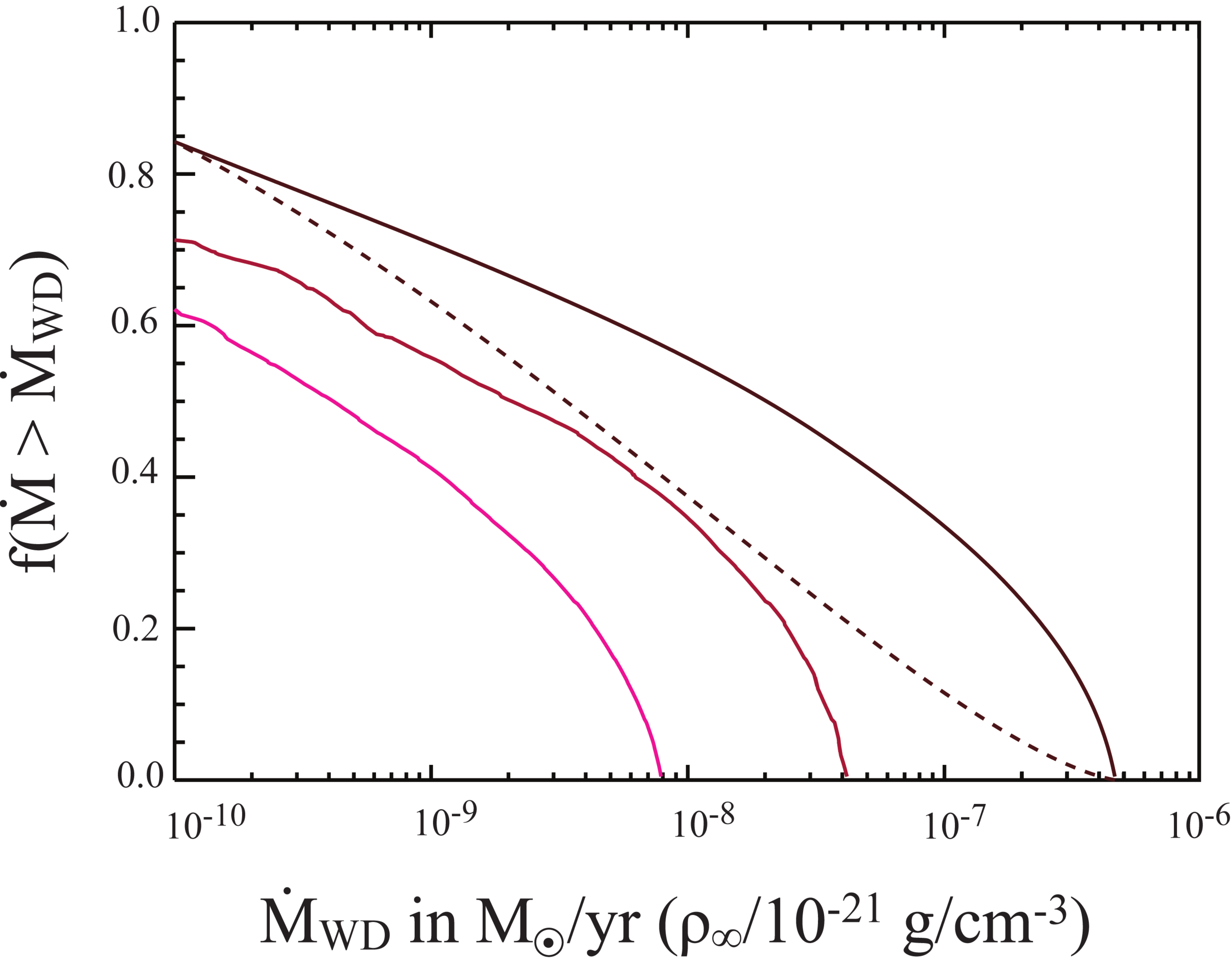}
\caption{The mass accretion rate fraction, $f$, of the
white dwarf cluster members  for both centrally condensed (solid line) and non-condensed (dashed line) compact remnant
distributions. The gas density profile  for model A has been used, with $\rho_\infty = 10^{-21} \, \rm{g \, cm^{-3}}$. To calculate the mass accretion distributions we have assumed
 $M_{\rm wd} = 1\,M_\odot$ and  $u_{\rm wd} = 15 \, \rm{km \, s^{-1}}$.
The red (brown)  line was calculating using the  numerical density  profile  for $\mu_\infty = 2.0\,(0.5)$. The black line assumes the analytic profile for $\mu_\infty = 2.0$.}
\label{fig:wdmdot}
\end{figure}

\end{document}